\documentstyle[12pt]{article}
\input epsf

\def\theequation{\arabic{section}.\arabic{equation}}
\def\ee{\end{equation}}
\def\be{\begin{equation}}
\newcounter{subequation}[equation]
\makeatletter
\@addtoreset{equation}{section}
\expandafter\let\expandafter\reset@font\csname reset@font\endcsname
  {\endeqnarray\stepcounter{equation}}
\makeatother
\normalbaselines 
\begin{document} 

\title{From Gauging Nonrelativistic Translations to N-Body Dynamics}

\author{J. Lukierski\thanks{e-mail: lukier@ift.uni.wroc.pl}
\\ \\
Institute for Theoretical Physics, University of Wroc\l aw, \\
 pl. Maxa Borna 9, 50-204 Wroc\l aw, Poland
\\ \\
P.C. Stichel\thanks{email: pstichel@gmx.de}
\\ \\
An der Krebskuhle 21\\ D-33619 Bielefeld\\ \\
W.J. Zakrzewski\thanks{e-mail: W.J.Zakrzewski@durham.ac.uk}
\\ \\
Department of Mathematical Sciences,University of Durham, \\
Durham DH1 3LE, UK}
\date{}
\maketitle
  \begin{abstract}
We consider the gauging of space translations with time--dependent
gauge functions. Using fixed time gauge of relativistic theory,
we consider the gauge--invariant model describing the motion of
nonrelativistic particles. When we use gauge--invariant
nonrelativistic velocities as independent variables the
translation gauge fields enter the equations through a $d\times
(d+1)$ matrix of vielbein fields and their Abelian field strengths,
which can be identified with
  the torsion tensors of teleparallel formulation of
relativity theory. We consider the planar case ($d=2$) in some
detail, with the assumption that the action for the dreibein fields is
given by the translational
  Chern--Simons term. 
We fix the asymptotic transformations in such a way that the space part of the 
metric becomes asymptotically Euclidean. The residual symmetries are 
(local in time) translations and rigid rotations. We describe the effective 
interaction of the $d=2$ $N$-particle problem and discuss its classical 
solution for $N=2$.
The phase space Hamiltonian $H$ describing two-body interactions satisfies a nonlinear equation $H={\cal H}(\vec x,\vec p;H)$ which implies, after
quantization, a nonstandard form of the Schr\"odinger equation with energy dependent fractional angular momentum eigenvalues. Quantum solutions of the two-body problem are discussed. The bound states with discrete energy levels
correspond to a confined classical motion (for the planar distance between two particles $r\le r_0$) and the scattering states with continuum energy 
correspond to the classical motion for $r>r_0$.
We extend our considerations by introducing an external constant magnetic 
field and, for $N=2$, provide the classical and quantum solutions in the 
confined and unconfined regimes.
\end{abstract}
\section{Introduction}

Our aim here is to discuss theories invariant under local
time--dependent nonrelativistic translations ($\vec{x} =
(x_{1}\ldots x_{d})$):
\begin{equation}
x^{\prime}_{i} = x^{\prime}_{i}(\vec{x}, t)
\end{equation}
supplemented by global space rotations ($x^{\prime}_{i}=R_{i}^{\
j} x_{j}$) and global time translations
\begin{equation}
t^{\prime} = t+a.
\end{equation}
The usual approach to local coordinate invariance in nonrelativistic
theory is to consider the limit $c\to \infty$ of a relativistic generally
 covariant theory. In particular it should be pointed out that
the nonrelativistic limit of the Einstein  action coupled to
a relativistic point particle had been explicitly performed by
Lusanna et al [1]. In this paper, following the earlier treatments by
one of the present authors in the case of one--dimensional model
[2,3], we shall impose nonrelativistic framework directly by 
 constructing actions covariant under (1.1--2).

We shall consider the problem here by using the vielbein formulation
of $(d+1)$--dimensional relativistic gravity, with $d$ vectors 
$E^{\underline{a}}_{\mu}$ ($\underline{a}=1,\ldots , d; \mu=0,1,\ldots d$) 
describing translational gauge fields\footnote{This idea goes
back to the papers by Cho [4], Hayashi and Shirafuji [5,6]; for
a review see [7]. Such framework leads to the so--called teleparallel
formulation of relativity (see e.g. [5--10]), with vanishing
curvature and nonvanishing torsion.}
 obtained from a relativistic $(d+1)\times (d+1)$ vielbein
  $ E^{\underline{\rho}} _{\mu} $  
   $(\rho = 0,1,\ldots d )$
  by fixing completely
  $E^{\underline{0}}_{\mu}$. We will find that in the equation of
motion for gauge--invariant nonrelativistic velocities
\begin{equation}
\xi^{\underline{a}} = 
E^{\underline{a}}_{\ i}\, \dot{x}_{i} +
E^{\underline{a}}_{0} \, ,
\end{equation}
the derivative of the vielbein field will enter only through its Abelian
field strength
\begin{equation}
T_{[\mu\nu]}^{\ a} =
\partial_{\mu}
E^{\underline{a}}_{\nu} - 
\partial_{\nu}
E^{\underline{a}}_{\mu}
\end{equation}
thus leading to the interpretation in terms of
components of the  torsion tensor.

The aim of this paper is to consider the dynamical consequences
of the coupling of nonrelativistic particles in two space
dimensions to nonstandard $D=2+1$ gravity action. 
It has been found in analogous one-dimensional model [2,3] that
gauging of nonrelativistic translations with Maxwell-like field
action quadratic in the torsion tensor leads to 
confinement via the geometric
bag formation. Further one can postulate\footnote{It has been
hypothesized already by Hehl that ``torsion should arise in the
microphysical realm" [11]. Some time ago [12] it was also shown 
that the gauge theory of gravity with torsion can lead to a
confining potential. }
 that such a
confinement can occur on a new level of microscopic theories of
 fundamental interactions and does not appear in macroscopic
``physical" gravity.
 The main result of the present paper 
 is to show\footnote{The preliminary results were presented in [13].}
   that the geometric bag solutions which may describe
confinement occur also in the planar case.

Our paper is organized as follows.
  Firstly, in Sect.~2 we shall consider $d$-dimensional
nonrelativistic point particles, by using the Lagrangian framework
covariant under (1.1). After considering in more detail the
 reparametrization--invariant
  nonrelativistic 
   particle dynamics for $d=1,2$ and $3$, in Sect. 3 we shall
consider the field actions for vielbein fields, using known 
results from the teleparallel formulation of relativistic gravity  in
$1+1$, $2+1$ and $3+1$ dimensions. 
We shall find that the $d=1$ 
action proposed in [2] is obtained from the 
 well--known  action in
$D=1 +1$ Einstein--Cartan gravity [14]; for $D=2+1$ we will
 consider the field action given by the so--called translational 
 Chern--Simons term [15-17]. 

In Sect. 4 we solve the field equations having chosen the
appropriate boundary conditions (gauge fixing). The general
considerations of the $N$-particle classical particle dynamics in
Sect. 5 is specialized in Sect. 6 to the two body problem.
Its explicit classical solution is presented there.
  Quantization of the two
body problem described by nonstandard Schr\"{o}dinger  equation
and its solutions, in particular
 numerical calculations of energy levels and wave
functions  in the confinement regime are given in Sect. 7.
The classical dynamics and quantized solutions in the presence of an external constant
magnetic field are described in Sect. 8. Sect. 9 reports some of
our conclusions and describes an outlook for further investigations.

\section{Nonrelativistic Particles and their Covariant Coupling
to Vielbein Fields}

Let us assume that $d$ space coordinates  $\vec{x}(t) = (x_{1}(t)
\ldots x_{d}(t))$ describe a trajectory of a $d$--dimensional
point particle with dynamics invariant under the transformations
(1.1). The velocities $\dot{x}_{l} \equiv {dx_{i}\over dt}$ then
transform under (1.1) as follows:
\begin{equation}
\dot{x}^{\prime}_{l}= {\partial x^{\prime}_{i}\over
\partial x_{j}}  
\, \dot{x}^{\prime}_{j}
+{\partial x^{\prime}_{i}\over
\partial t}.
\end{equation}
The formula (2.1) can be obtained from the $(d+1)$--dimensional
``relativistic" formula ($x_{\mu}=(\vec{x},x_{0})$)
\begin{equation}
{\dot{x}}^{\prime\mu} = 
{\partial x^{\prime\mu}\over
\partial x^{\nu} }\, {\dot{x}}^{\nu}
\end{equation}
if we assume that $x^{0}=-x_{0} = -t$, {i.e.}\footnote{The
condition (2.3) can be treated as fixing the general
reparametrization 
 $x^{\prime}_{0}=x^{\prime}_{0} (\vec{x},x_{0})$
 of $(d+1)$-th coordinate.}
 \begin{equation}
{ \dot{x}}_{0}=1.
 \end{equation}
 The preservation of (2.3) in any coordinate frame implies that
 \begin{equation}
 {\partial x^{\prime}_{0}\over
\partial x_{i} } = 0
\qquad
 {\partial x^{\prime}_{0}\over
\partial t}= 1  \Rightarrow x^{\prime}_{0} = x_{0} +a 
 \end{equation}
 in accordance with (1.1).
 
 It is well--known how to introduce the compensating gauge fields
for the transformations (2.2); namely, we should 
replace the velocities by the world scalars ({i.e.} scalars
under the transformations (2.2)) which are vectors in tangent space
\begin{equation}
{\dot{x}}^{\mu} \to
\xi^{\underline{\mu}} 
= E^{\underline{\mu}}_{\ \nu} 
{\dot{x}}^{\nu}
=\left( \xi^{\underline{0}}, 
\xi^{\underline{a}} \right).
\end{equation}
Here the $(d+1)\times (d+1)$--bein 
$E^{\ \underline{\mu} }_{   \nu }$
transforms as a covariant vector under the local transformations (1.1)

\begin{equation}
E^{\prime \underline{\mu} }_{ \nu }
=
 {\partial x^{\rho}\over
\partial x^{\prime \, \nu} } \,
E^{\underline{\mu} }_{\rho}
 \end{equation}
 and as a global $(d+1)$--dimensional vector under the Lorentz
rotations in the tangent space. The imposition of relations
(2.3--4) and their validity in any coordinate frame
  imply that one should choose
\begin{equation}
E^{\underline{0}}_{\ 0} = 1  \qquad 
E^{\underline{0}}_{\ {i}} = 0
\end{equation}
as then one gets $\xi^{\underline{0}} = \dot{x}^{0} = -1$. The choice
(2.7) we shall call the nonrelativistic gauge because it splits the
$(d+1)$--dimensional Lorentz vector into a $d$--dimensional
nonrelativistic vector and a scalar as well as because  it implies the
Newtonian notion of absolute time. One can write $(a,b=1, \ldots
d)$
\begin{equation}
E^{\underline{\mu} }_{\ \nu }=
\pmatrix{
\matrix{1,} &\matrix{0, &\ldots &0}
\cr\cr
\matrix{ e^{\underline{1}} \cr \vdots \cr e^{\underline{d}} } 
&\matrix{h^{\underline{a}}_{\ i} }
}
\end{equation}
and introduce the inverse vielbein 
${{E}}^{\nu}
_{\ \underline{\mu}}$ as follows:
\begin{equation}
E_{\underline{\mu}}^{\ \nu }=
\pmatrix{
\matrix{1,} &\matrix{0, &\ldots &0}
\cr\cr
\matrix{{e}^{{1}}\cr \vdots \cr{e}^{{d}}} 
&\matrix{{h}^{i}}_{\underline{a} }
},
\end{equation}
where $h^{\underline{a}}_{\ i}{h}^{i}_{\ \underline{b}}
=\delta^{\underline{a}}_{\ \underline{b}}$
and $e^{i}$, $e^{\underline{a}}$ are independent.

We replace the ``flat" free nonrelativistic Lagrangian by
\footnote{The $d$--dimensional indices, due to the Euclidean
nonrelativistic metric, can be taken equivalently as lower or
upper indices.}
\begin{equation}
{\cal L}_{0} = {m\over 2} \dot{\vec{x}}\, ^{2} \Rightarrow 
{\cal L}_{0} + {\cal L}_{mt} = {m\over 2}
\xi_{\underline{a}} \, \xi^{\underline{a}}\, .
\end{equation}

Using formula (2.5) for the nonrelativistic covariantized
velocity one obtains the following Euler--Lagrange equations of motion
\begin{equation}
h^{\underline{a}}_{\ i}
\, \dot{\xi}_{\underline{a}} - T^{\underline{a}}_{i0} \, \xi_{\underline{a}} =
T^{\underline{a}}_{ij} \, \xi_{\underline{a}}\, {\dot{x} j}
\end{equation}
or, using (1.3) and then
${{E}}^{i}
_{\ \underline{\mu} } 
{E}^{\mu}_{\ 0}
=0 $, one gets 
$\dot{x}^{j}={h}^{j}_{\underline{b}}
\xi^{\underline{b}}- {e}^{j}$ and
\begin{equation}
\dot{\xi}_{\underline{c}} - {h}^{i}_{\ \underline{c}}
\, {h}^{j}_{\ \underline{b}}\,
T^{\underline{a}}_{ij}\,  \xi_{\underline{a}}\, \xi^{\underline{b}} -
{h}^{i}_{\ \underline{c}} \,
T^{\underline{a}}_{i0}\, \xi_{{a}}=0,
\end{equation}
where the tensors $T^{\underline{a}}_{ij}$, $T^{\underline{a}}_{i0}$ are given
by the formulae (see also (1.4))
\begin{equation}
T^{{a}}_{ij} =h^{\underline{a}}_{i,j} - h^{\underline{a}}_{j,i} \qquad 
T^{\underline{a}}_{i0} = h^{\underline{a}}_{i,0} - e^{a}_{, i}\, .
\end{equation}

In particular, for  $ d=1$ we have only two zweibein--components 
$E^{\underline{1}}_{0} =e$, 
$E^{\underline{1}}_{1} =h$ and $T^{\underline{1}}_{ij}=0$,
$T^{\underline{1}}_{i0}=\partial_{t}h -\partial_{x}e$ (see [2]). In
$d=2$ we have
\begin{equation}
h^{\underline{a}}_{i} \ = \ 
\pmatrix{ h^{\underline{1}}_{1}, &h^{\underline{1}}_{2}
\cr\cr
h^{\underline{2}}_{1}, & h^{\underline{2}}_{2}
}
\qquad \qquad e^{\underline{a}} = (e^{\underline{1}}, e^{\underline{2}})
\end{equation}
and it is easy to see that the term 
\begin{eqnarray*}
T^{\underline{a}}
_{ij} \, 
\xi_{\underline{a}} \,
\dot{x}^{j} = 
\varepsilon_{ij}
T^{a}\,
 \xi_{\underline{a}}\,
 \dot{x}^{j} \qquad (T^{a}={1\over 2} \varepsilon^{ij}
T^{a}_{ij})
\end{eqnarray*}
does not vanish.

One can also write Lagrangian (2.10) for any $d$ as 
\begin{equation}
{\cal L}^{\rm cov}_{0} = {m \over 2}
\left(g_{ik} \, \dot{x}^{i}\, \dot{x}^{k}
+ 2 g_{i0} \, \dot{x}^{i} + g_{00}\right),
\end{equation}
where
\begin{equation}
g_{ik} = h^{\underline{a}}_{i}\,  h^{\underline{a}}_{k}
\qquad
g_{i0} = e^{\underline{a}}\,  h^{\underline{a}}_{i}
\qquad g_{00} = e^{\underline{a}} \, e^{\underline{a}} - 1.
\end{equation}

We see that one can treat the $g_{i0}$ components of the metric as
describing the magnetic field coupled to the velocity of
the particle\footnote{It is interesting to relate this observation
to the Kaluza--Klein description of the electric charge in terms of
the additional fifth dimension.} [18,19], and that the component $g_{00}$ 
describes a potential. The Euler--Lagrange equations of motion now take the
form\footnote{The equations (2.16) appeared also in
[1], as the nonrelativistic limit of a covariant relativistic
particle model.}

\begin{equation}
\ddot{x}^{i} +\Gamma^{i}_{jk} \, \dot {x}^{j}\, \dot{x}^{k}
= - g^{ij}\, {\partial g_{j}^{\ l }\over \partial t}
\, \dot{x}^{l}
 + g^ {ij} \, F_{j l} \, \dot{x}^{l} +
 F^{i} \left[ g_{ij}, g_{0i}, g _{00} \right],
 \end{equation}
 where $\Gamma_{jk}^{ \ l}$ is the symmetric Levi--Civita connection
 \begin{equation}
 \Gamma_{jk}^{ \ l} = {1\over 2}
 \left( {\partial g_{lj} \over \partial x_{k}} +
 {\partial g_{lk} \over \partial x_{j}} -
 {\partial g_{jk} \over \partial x_{i}}\right) \, ,
 \end{equation}
 the field $F_{jl}$ plays the role of the magnetic field strength ($F_{jl}
 = \partial_{j}A_{l} -\partial_{l}A_{j}$, with $A_{j}= g_{j0}$)
and the force $F^{i}$ is given by the formula
\begin{equation}
F^{i}= g^{ij}\, E_{j},
\end{equation}
where $E_{j}$ plays the role of electric field strength
\begin{equation}
E_{j} = {\partial A_{0} \over \partial x_{j} }-
{\partial A_{j} \over \partial t },
\qquad A _{0} = g _{00}.
\end{equation}
The most convenient form of the equations of motion depends on the
type of field action for the vielbein field. The formulation with basic 
variables
$\xi_{\underline{a}}$ is the best suited for the discussion of the
interactions with a gravitational field described by an action
functional which depends  only on vielbeins and on the torsion
variables (1.4).

\section{Torsion Lagrangians and Coupled Particle--Torsion
Field Systems}

Following M\"{o}ller [20] one can  postulate
vielbeins as fundamental variables in gravity theory and
treat the metric as a derived quantity.
Let us recall here that the covariantization of the Dirac equation
\underline{requires} vielbeins.
Moreover, tetrads
 appear naturally in the geometric framework as translational gauge fields,
providing the so--called teleparallel formulation of relativity
theory [5--10].

Because eq. (2.11--12) involve only tetrads and  torsion tensors, it is
convenient to consider the free gravity actions which depend only on
these field variables. In order to obtain nonrelativistic actions
we can consider the relativistic $D=d+1$--dimensional torsion
actions and set the time component $T^{0}_{\mu\nu}=0$, 
consistently with our nonrelativistic gauge condition (2.7).
Because the explicit forms of the torsion actions depend strongly on
dimensions, we shall consider below, separately, the three cases of $d=1,2$ and
$3$.

i) $d=1$

It is known that in $1+1$ dimension the Einstein--Hilbert action
is a topological invariant and so we can consider the  action
quadratic in curvature (see e.g. [21]) or in torsion [14]. The
quadratic torsion action has the form
\begin{equation}
\mbox{\LARGE\sl S}^{\ d=1}_{\ rT} = {1\over 2 \lambda }
\int dt \, dx\cdot \det E \cdot T^{\underline{a}}_{\mu\nu}\,
T_{\underline{a}}^{\mu\nu}
\qquad \cases{a=0,1 & \cr\mu,\nu=0,1 &}
\end{equation}
and  reduces to the nonrelativistic field action considered in [2,3]
when we observe that in the nonrelativistic gauge    ( 2.7)
\begin{equation}
\det E =  h \, .
\end{equation}
Putting $T^{\underline{0}}_{\mu\nu}=0$ and writing $F\equiv
{1\over h} T^{\underline{1}}_{\mu\nu}$ 
we get\footnote{See  formula (18) in [2].}
\begin{equation}
\mbox{\LARGE\sl S}^{\ d=1}_{\ nrT} 
= {1\over 2\lambda } \int dt \, dx \,
h\cdot F^{2}.
\end{equation}

ii) $d=2$

The $(2+1)$--dimensional gravity with Hilbert--Einstein action
is dynamically trivial as outside nonvanishing matter
  sources the space--time is
flat (see e.g.  [22]). This fact was one of the reasons why the
alternative 3--dimensional topological gravity models -- without
torsion [23] and with torsion  [16,17] had been considered. In
particular, one can introduce as a candidate for a $D =2+1 $ gravity
action the translational Chern--Simons term [16,17] ($t=x_{0};
\mu,\nu,\rho= 0,1,2,  \underline{\alpha},  \underline{\beta}=0,1,2$;
 $\eta_{\alpha \beta} = {\rm diag} (-1,1,1)$)
\begin{equation}
\mbox{\LARGE\sl S}^{\ d=2}_{\ nrT} 
= {1\over \lambda} \int d^{3}x \,
\varepsilon^{\mu\nu\rho} \, E^{\underline{\alpha}}_{\mu} \,
T^{\underline{\beta}}_{\nu\rho} \, \eta_{\alpha\beta}.
\end{equation}
If we pass to the nonrelativistic gauge (2.7) we obtain
($\underline{a}, \underline{b} = 1,2$)

\begin{equation}
\mbox{\LARGE\sl S}^{\ d=2}_{\ nrT} 
= {1\over \lambda} \int dt\,  d^{2}x \,
\varepsilon^{\mu\nu\rho} \, E^{\underline{a}}_{\mu} \,
T^{\underline{a}}_{\nu\rho}. 
\end{equation}
The action (3.4) will be used in the next Section  to derive and study
various properties of the
planar $N$--body interactions. 
 
 As later on we will study in detail the $d=2$ case let us
describe in some detail the planar coupled system of equations
describing an interacting particle -- torsion field system, with
torsion fields described by the Abelian Chern--Simons action (3.5).
Introducing $N$ trajectories $\vec{x}_{\alpha}(t) = x^{i}_{\alpha}(t)$
 ($i=1,2; \alpha=1,\ldots N$) for $N$ particles and the notation
 \begin{equation}
 E^{\underline{a}}_{\mu; \alpha}(t) \equiv
 E^{\underline{a}}_{\mu}(\vec{x}_{\alpha}(t), t),
\end{equation}
the free action for $N$ particles in $d=2$ dimensions, in 
the first order formalism,
can be written as 
\begin{equation}
\mbox{\LARGE\sl S}^{\ (N)}_{\ \rm part}= -
\sum\limits^{N}_{\alpha = 1} m_{\alpha}
\int dt \left[ {1\over 2} \xi^{\underline{a}}_{\alpha}
\, \xi^{\underline{a}}_{\alpha} -
\xi^{\underline{a}}_{\alpha}\left(
h^{\underline{a}}_{j; \alpha}\, \dot{x}^{j}_{\alpha} +
e^{\underline{a}}_{\alpha} \right)\right],
\end{equation}
thus providing the constraint formula
\begin{equation}
\xi^{\underline{a}}_{\alpha} = h^{\underline{a}}_{j; \alpha}
\dot{x}^{j} + e^{\underline{a}}_{\alpha}.
\end{equation}
The action (3.7) can be written in the field form, using the
mass density function, as 
\begin{equation}
\mbox{\LARGE\sl S}^{\ (N)}_{\ \rm part}= - \int dt d^{2}x \left[
 {1\over 2} \xi^{\underline{a}}
\, \xi^{\underline{a}} -
\xi^{\underline{a}}\left(
h^{\underline{a}}_{j}\, \dot{x}^{j} +
e^{\underline{a}} \right)\right]
\rho_{r}(\vec{x};\vec{x}_{1}(t)\ldots \vec{x}_{N}(t)),
\end{equation}
where 
\begin{equation}
\rho_{N}(\vec{x};\vec{x}_{1}(t)\ldots \vec{x}_{N}(t))=
\sum\limits^{N}_{\alpha = 1} m_{\alpha}  \, \delta^{(2)} 
\left(\vec{x} - \vec{x}_{\alpha}(t)\right).
\end{equation}
The field action (3.5) thus takes the form of the field action 
($i,j,k=1,2$)
\begin{equation}
\mbox{\sl S}_{\ \rm field}= {1 \over \lambda} \int dt d^{2}x
\left( e^{\underline{a}}  B^{\underline{a}} -
\varepsilon_{jk} \, h^{\underline{q}}_{j} 
{\cal E}^{\underline{a}}_{k}  \right),
\end{equation}
where
\begin{equation}
{B}^{\underline{ a}} =
\varepsilon_{jk} \, \partial_{j} \, h^{\underline{a}}_{k}
\qquad
{\cal E}^{\underline{ a}}_{k} =
 \partial_{t} \, h^{\underline{a}}_{k}-
  \partial_{k} \, e^{\underline{a}}.
  \end{equation}
  The fields ${B}^{\underline{ a}}$ and ${\cal E}^{\underline{
a}}_{k}$ plays the role of the magnetic and electric fields, respectively, 
with the internal $O(2)$ index $\underline{a}$ describing the $d=2$
nonrelativistic rotation group\footnote{In the vielbein formalism
the  rotation group (Lorentz in the relativistic case, Euclidean in
the nonrelativistic gauge) plays the role of an internal symmetry.}.

We can now derive the coupled equations describing the 
 $d=2$ particle--field system, 
 described by the action $S =\mbox{\sl S}^{\ (N)}_{\ \rm part}+
\mbox{\sl S}_{\ \rm field}$ (see    (3.7) and (3.9)). The
equations for the particle trajectories (see (2.11)), having used the
notation (3.12), then take the form:
\begin{equation}
h^{\underline{a}}_{\ i} \, \dot{\xi}^{\underline{a}}_{\alpha}
-\varepsilon_{ij}\, B^{\underline{a}}_{\alpha}\,
\xi^{\underline{a}}_{\alpha}\, \dot{x}^{j} -
 {\cal E}^{\underline{a}}_{j;\alpha}\,
\xi^{\underline{a}}_{\alpha}= 0.
 \end{equation}
 We can now pass to the form (2.11) and  use the following explicit
form for the  nonrelativistic dreibein (2.9)
\begin{equation}
 {E}^{\nu}_{\ \mu}= {1\over d}
 \pmatrix{d & 0 & 0 \cr\cr
 e^{\underline{2}} h^{\underline{1}}_{\ 2}
  - e^{\underline{1}} h^{\underline{2}}_{\ 2},
 & h^{\underline{2}}_{\ 2}, & -h^{\underline{1}}_{\ 2},
 \cr\cr
  e^{\underline{1}} h^{\underline{1}}_{\ 1}
   -  e^{\underline{2}} h^{\underline{1}}_{\ 1},
   &- h^{\underline{2}}_{\ 1},
   & h^{\underline{1}}_{\ 1}
   },
   \end{equation} 
where $d=\det ( h^{\underline{l}}_{\ j})= h^{\underline{1}}_{\ 1}
 h^{\underline{2}}_{\ 2}-  h^{\underline{1}}_{\ 2} h^{\underline{2}}_{\ 1}$.

The field equations for the dreibein fields then take the form:
\setcounter{equation}{0}
\renewcommand{\theequation}{\thesection.15\alph{equation}}

\begin{equation}
{\cal E}^{\underline{a}}_{k}(\vec{x},t)= -
{\lambda \over 2} \sum\limits_{\alpha} m_{\alpha}
\varepsilon_{kj}
\,  \dot{x}^{j}_{\alpha}  \cdot
\xi^{\underline{a}}_{\alpha}\cdot
\delta^{(2)} (\vec{x}-\vec{x}_{\alpha}(t)),
\end{equation}

\begin{equation}
{ B}^{\underline{a}}(\vec{x},t)= -
{\lambda \over 2} \sum\limits_{\alpha} m_{\alpha}
\xi^{\underline{a}}_{\alpha}
\delta^{(2)} (\vec{x}-\vec{x}_{\alpha}(t)).
\end{equation}

We see, therefore, that the dreibein field equations,
 consistently with the 
 general property of the $d=2+1$
gravity, imply that the space--time is flat outside of the matter
sources. 

In the next section we shall 
try to solve the system of equations (3.13) and (3.15a-b) by 
  using the technique of singular gauge 
transformations (see e.g. [24--28]).

iii) $d=3$

In $3+1$ dimensions we can write three independent quadratic torsion
actions [5--10]. It is interesting to observe that there are
special linear combinations of quadratic torsion terms which
provide four--dimensional Hilbert--Einstein action. The detailed
consideration of coupled nonrelativistic particle -- torsion
fields systems\footnote{In our framework, due to the relation
(1.4), the local frame fields $E^{a}_{\mu}$ can be called torsion potentials.}
 is postponed to a future investigation.
\renewcommand{\theequation}{\thesection.\arabic{equation}}

\section{Solution of the Field Equations}
\subsection{Gauge fixing and residual symmetry}

\renewcommand{\theequation}{\thesection.\arabic{equation}}

The set of equations (3.15a-b) can be rewritten as (see (3.5)):
\begin{equation}
\epsilon^{\mu\nu\rho}   \partial_{\mu}
E^{\underline{a}}_{\nu} = 
-{\lambda \over 2} \sum\limits_{\alpha} \xi^{\underline{a}}_{\alpha}
\,           \dot{x}^{\rho,\alpha}
\, \delta(\underline{x}- x^{\alpha}),
\end{equation}
where (3.15a) and (3.15b) are obtained respectively by putting
 $\mu=1,2$ ($\mu\equiv \alpha$) and $\mu=0$.
 
 The general solution of (4.1) can be written in the pure gauge
 form
 
 \renewcommand{\theequation}{\thesection.2\alph{equation}}
 \setcounter{equation}{0}
 \begin{equation}
E^{\underline{a}}_{\mu}(\vec{x},t) = 
\widetilde{E}\ ^{\underline{a}}_{\mu}(\vec{x},t)
 +  
 \partial_{\mu} \Lambda^{\underline{a}}(\vec{x},t) 
 = 
  \partial_{\mu} \tilde{\Lambda}\,^{\underline{a}}(\vec{x},t) 
 \end{equation}
 where
   
    \begin{equation}
\widetilde{E}\ ^{\underline{a}}_{\mu}(\vec{x},t) = 
-{\lambda \over 4\pi} \partial_{\mu}
 \sum\limits_{\alpha} \xi^{\underline{a}}_{\alpha}
\,\Phi(\vec{x}- \vec{x}_{\alpha})
\end{equation}
\renewcommand{\theequation}{\thesection.\arabic{equation}}
\setcounter{equation}{2}
and

   - $\Lambda^{\underline{a}}$ is an $O(2)$--vector valued pair
of regular gauge functions,

- $\Phi(\vec{x})$ is a singular gauge function satisfying the
following equation (see e.g. [26--28])
 \begin{equation}
\epsilon^{ij}   \partial_{i}
\partial_{j}
\Phi(\vec{x}) = 2\pi \, \delta(\vec{x}\, )
 \end{equation}
thus expressing the singular nature of the first term in (4.2b).
As a solution of (4.3) we can
take
\begin{equation}
\Phi(\vec{x}\, ) = \mbox{arc}\,\mbox{tan} {x_{2}\over x_{1}}
\end{equation}
{i.e.}
\begin{equation}
\partial_{k}\Phi({\vec{x}})  =- \varepsilon_{kl}\partial_{l}\ln
|\vec{x}\, |.
\end{equation}
The function  $\partial_k\Phi(\vec{x})$ can be regularized in such a way
that it has a well defined 
limit for $\vec{x}\rightarrow 0$, {e.g.} we can replace
  $\ln|\vec{x}\, |$ in (4.5) by [28]
\begin{equation}
\ln   | \vec{x} \, | \longrightarrow
\ln^{(\epsilon)}  \left(      \vec{x}  \right) : = \,
{ 1\over \epsilon \pi} \int d^{2} y \, \ln | \vec{x} - \vec{y}\, |
e^ {- {y^{2}\over \epsilon}}.
\end{equation}
In this case we find
\begin{eqnarray}
&{i)}\qquad \qquad\qquad &\ln^{(\epsilon)}
 (\vec{x})  \stackrel{\epsilon \searrow   0}{\longrightarrow}
\ln |\vec{x} |,
\cr\cr
 &{ii)}\qquad \qquad \qquad &\lim\limits_{\vec{x} \to 0} 
 \varepsilon_{kl}\partial _{l} \ln ^{(\epsilon)}(\vec{x})
 \longrightarrow 0 \qquad \qquad  \forall \varepsilon>0.
 \end{eqnarray}

Let us note that the solutions for the fields 
$E_\mu\sp{\underline{a}}(\vec{x},t)$ with asymptotically
 nonvanishing gauge function $ \Lambda^{\underline{a}} $
 do not solve  the Hamilton's variational principle for the
field action (3.11).
The bad asymptotic  behaviour
for $r\rightarrow \infty$
 of dreibeins $ E_\mu\sp{\underline{a}}(\vec{x},t)
$ leads to the appearence of  nonvanishing  surface
integrals and, in consequence, our $ E_\mu\sp{\underline{a}}(\vec{x},t)$ 
do not minimize the action.
Such a situation arises also in General Relativity     (see e.g. [29]).
In the following
we will argue along similar lines  by choosing 
an appropriate asymptotic form for $E_\mu\sp{\underline{a}}$ 
and adding two surface integrals to the action.

To do this we decompose
\begin{equation}
E_\mu\sp{\underline{a}}(\vec{x},t)\,=\tilde E_\mu\sp{\underline{a}}(\vec{x},t)
\,+\, E_\mu\sp{ {\rm as}\,  \underline{a}}
\label{decomp}
\end{equation}
where we require that the new field variables 
$ \tilde E_\mu\sp{\underline{a}}(\vec{x},t)$ satisfy 
\begin{equation}
\tilde E_\mu\sp{\underline{a}}(\vec{x},t)\,\rightarrow \,O(r\sp{-1})
\quad \mbox{as}\quad r\rightarrow\infty
\label{asymp}
\end{equation}
and  we assume that the
 asymptotic parts $ E_\mu\sp{as \underline{a}}$ are given as functions of $t$ 
only.
Then from the 
 pure gauge form
\begin{equation}
E_\mu\sp{as\underline{a}}(t)=\partial_{\mu} 
\Lambda\sp{\underline{a}}(\vec{x},t).
\label{4.11}
\end{equation}
we obtain
\begin{equation}
  \Lambda^{i}(\vec{x},t) =    x^ {i} -  a^{i}(t)\label{gauge}
\end{equation}
giving us
\begin{equation}
E_0\sp{as \underline{a}}\,=\,-\dot a\sp{\underline{a}}(t)
\,=:\,-v\sp{\underline{a}}(t),\qquad E_i\sp{as \underline{a}}\,=\,
\delta_i\sp{\underline{a}},\label{para}
\end{equation}
where we have chosen the factor in front of $x\sp{i}$ to be equal 
to one in order to have, asymptotically, the Euclidean metric.

 The choice (4.11) for $\Lambda^{i}$ breaks asymptotically
the invariance with respect to local space translations
(1.1). In the formula (4.2a) only the fields  $\tilde E_\mu\sp{\underline{a}}$
 transform covariantly
with respect to those local translations which preserve the
asymptotic behaviour (4.9). 
However, as under general coordinate transformations the functions
 $\widetilde{\Lambda}\,^{i}$ are scalars, the changes 
  $\delta \, \widetilde{\Lambda}\, ^{i}$
  under (1.1) of both the singular and regular parts of 
  $\widetilde{\Lambda}\,^{i}$ must separately vanish and we obtain
    \begin{equation}
\delta x\sp{i}\,=\,\delta a\sp{i}(t),
\end{equation} 
where $x^i$ and  $a^i$ transform as vectors under rotatons in
tangent space. Therefore, as a residual symmetry, we obtain
translations, local in time, and rigid rotations (see also [30-32]
for a similar situation in General Relativity).

Putting (4.8) and (4.11) into the Chern--Simons 
action (see (3.5) and (3.11))  we find (modulo a total
time derivative)
\begin{equation}
L_{field}\,=\,\tilde L_{field}\,-\,I_1\,-\,I_2
\end{equation}
where
\begin{equation}
\tilde L_{field}:=\,L_{field}[\tilde E_\mu\sp{\underline{a}}]
\end{equation}
and the integrals $I_{1,2}$ are defined by

\begin{eqnarray}
I_1\,=&\,{1\over \lambda}\,\int\,d\sp2x\, 
v\sp{\underline{a}}\cdot \epsilon\sp{ij}
\,\partial_i\,\tilde
E_j\sp{\underline{a}}(\vec{x},t)\\
I_2\,=&\,{1\over \lambda}\,\int\,d\sp2x\, 
\epsilon\sp{ij}
\,\partial_i\,\tilde
E_0\sp{\underline{j}}.
\end{eqnarray}

Next we use the Stokes' theorem to rewrite $I_{1,2}$ as
\begin{eqnarray}
I_1\,=&\,{1\over \lambda}\,\lim_{r\rightarrow\infty}\,r v\sp{\underline{a}}
\int_0\sp{2\pi}\,d\varphi\,  \left(-\tilde E_1\sp{\underline{a}}
\,sin(\varphi)\,+\,
\tilde E_2\sp{\underline{a}}\,cos(\varphi)\right),\\
I_2\,=&\,{1\over \lambda}\,\lim_{r\rightarrow\infty}\,r 
\int_0\sp{2\pi}\,d\varphi\, \left(-\tilde E_0\sp{1}\,sin(\varphi)\,+\,
\tilde E_0\sp{2}\,cos(\varphi)\right).
\end{eqnarray}

Given the asymptotic behaviour (4.9) the boundary integrals $I_{1,2}$  
exist but they do not vanish, in general, and so only the
modified field Lagrangian
\begin{equation}
\tilde{L}_{field}\,=\,L_{field}\,+\,I_1\,+\,I_2
\end{equation}
has well defined functional derivatives 
with respect to the field $\tilde E_\mu\sp{\underline{a}}$.

Due to the asymptotic behaviour (4.9) the variations of 
 $\tilde E_\mu\sp{\underline{a}}$ and $v^{\underline{a}}$  
  are independent of each other {\it i.e.} the fields
   $\tilde E_\mu\sp{\underline{a}}$
  and $v^{\underline{a}}$ appear as new variables. The property that  in
a $(2+1)$-dimensional gravity boundary terms give rise to
additional degrees of freedom has been shown also in [32].
 
 The particle action now takes the form
 \renewcommand{\theequation}{\thesection.21\alph{equation}}
 \setcounter{equation}{0}
 \begin{equation}
S^{(N)}_{\rm part}[x,\xi,E]\,=\,S^{(N)}_{\rm part}[x,\xi,\tilde E]\,
+\, \int \,dt\,\left[\,
\sum\limits_{\alpha}\xi_{\alpha}\sp{\underline{a}}
\cdot \dot x_{\alpha}\sp{\underline{a}}\,-\,
v\sp{\underline{a}}\,\cdot\,\sum\limits_{\alpha}
\xi_{\alpha}\sp{\underline{a}}\right].
\end{equation}
and the modified field action 
$\tilde{S}_{\rm field} =S_{\rm field}[\widetilde{E}]$ 
(see (4.20))  
 is given by $S^{(N \, R)}_{T}$ (eq. 3.5) but taken as a function
of $\tilde{E}$
\begin{equation}
S_{\rm field}[\widetilde{E}] =  S^{(N\, R)}_{T}[\widetilde{E}]
 \end{equation}
 \renewcommand{\theequation}{\thesection.\arabic{equation}}
 \setcounter{equation}{21}
 We see that in the action (4.21) there are separated the
variables describing the ``bulk" $(\widetilde{E}\,
_{\mu}^{\underline{a}})$ and asymptotic behaviour $(v^a)$.

The new variables $(v^{\underline{a}})$ describing local gauge
degrees of freedom appear as the Lagrange multipliers in (4.21a)
 and are not determined by the EOM (cp. [33]). By fixing it as a
constant we obtain Galilei invariance as the residual symmetry.

If we vary  $\tilde E_\mu\sp{\underline{a}}$
we obtain  the field
equations (4.1) whose solutions should be taken, 
 in accordance with (4.9), as
\begin{equation}
\tilde E_\mu\sp{\underline{a}}(\vec{x},t)\,=\,-{\lambda\over 4\pi}
\,\partial_{\mu}\sum\limits_{\alpha} \xi^{\underline{a}}_{\alpha}(t)
\Phi(\vec{x}-\vec{x}_{\alpha}(t)).
\end{equation}

By varying $S$ with respect to $v\sp{\underline{a}}$ we obtain the
constraint
\begin{equation}
\sum\limits_{\alpha}\xi_{\alpha}\sp{\underline{a}}\,=\,0
\label{constraint}.
\end{equation}

We note that only the presence of the constraint (4.23) yields 
the correct asymptotic behaviour 
 as well as the correct rotational properties 
of $\tilde E_0\sp{\underline{a}}$ because
\begin{itemize}
\item{}
from (4.22) we have
\begin{equation}
\tilde E_0\sp{\underline{a}}(\vec x,t)\,\rightarrow \,
-{\lambda\over 4\pi}\sum_{\alpha}\dot \xi_{\alpha}\sp{\underline{a}}\,
\Phi(\vec x)\,+\,O(r\sp{-1})\, ,
 \quad \mbox{as}\quad r\rightarrow \infty\
\end{equation}
which agrees with (4.9) if (4.23) holds.
\item{}
for a rotation by an angle
$\varphi$ in the 1-2 plane we have
\begin{equation}
\Phi\rightarrow \Phi+\varphi
\end{equation}
and so with (4.24) we get
\begin{equation}
\tilde E_0\sp{\underline{a}}\,\rightarrow\,\tilde E_0\sp{\underline{a}}\,
-\, 
 \Phi \, 
{\lambda\over 4\pi}\sum_{\alpha}\dot \xi_{\alpha}\sp{\underline{a}},
\end{equation}
which agrees with (2.6) only if (4.23) holds.
\end{itemize}

 \subsection{Conservation Laws}

Due to the constraint (4.23) 
 the total momentum of the system
 vanish.
 To see this let us note that after gauge fixing our 
Lagrangian $L$ is invariant
with respect to 
space translations local in time, and rigid  rotations.
According to Noether's theorem, 
the invariance of $L$ with respect to $\delta \vec{x}$ leads to a 
conserved quantity $C[\delta\vec{x}]$ 
 (see [34],  eq. (6.29c) for the corresponding case of the 
  Chern--Simons electrodynamics):
\begin{equation}
C[\delta\vec{x}]\,=\,\sum_{\alpha} 
p_{\alpha}\sp{\underline{a}}\cdot \delta x_{\alpha}\sp{\underline{a}}\,+\,
{2\over \lambda}\int\,d\sp2x\,B\sp{\underline{a}}\tilde{E}_k\sp{\underline{a}}
\,\delta x_k\,=\,C_{part}\,+\,C_{field}.
\label{eq1}
\end{equation}
Inserting  into (4.27) 
 the expression (3.15b)   for $B\sp{\underline{a}}$ 
and (4.22)  for 
$\tilde{E}_{k}\, ^{\underline{a}}$ 
   and taking into account 
  the definition of the canonical particle momentum
$p_{\alpha}\sp{\underline{a}}={\partial L\over \partial \dot x_{\alpha}\sp{\underline {a}}}$ we get
\begin{equation}
C[\delta\vec{x}]\,=\,\sum_{\alpha}\,
\xi_{\alpha}\sp{\underline{a}}\,\delta x_{\alpha}\sp{\underline
{a}}
\label{eq2}
\end{equation}
Taking $\delta \vec{x}$
 as describing respectively  space translations
 $(\delta{x}\,^{\underline{a}}_{\alpha} =
 \delta a \,^{\underline{a}})$
  and rotations
   $(\delta{x}\,^{\underline{a}}_{\alpha} =
 \varepsilon  \,^{\underline{a}\underline{b}} \, 
  x_{\alpha}^{\underline{b}}
  \cdot \delta \alpha)$
and denoting the corresponding conserved quantities by $P_i$ and $J$,
we obtain due to (3.15b) and (4.22) for $P_i$ the formula
\begin{equation}
P^i\,= P_{\rm part}^{i}   = \sum\limits_{\alpha} \xi^{i}_{\alpha}
\end{equation}
and for $J$ we get
\begin{equation}
J\,=\,\sum_{\alpha}\,\epsilon_{ij}\,x_{i,\alpha}\,\xi_{j,\alpha}.
\end{equation}
Observe that $J$ is different from the total canonical 
particle angular momentum
$J_{part}$. The field contribution $J_{field}$, which is obtained from
(3.15b) (4.22) and (4.23), namely,
\begin{equation}
J_{field}\,=\,-{\lambda\over 8\pi}\,
\sum_{\alpha}
 \xi_{\alpha}\sp{i}\,\xi_{\alpha}\sp{i}
\end{equation}
is conserved separately because it is proportional to the $N$-particle
Hamiltonian to be given in section 5.2.

The fact that $J\ne J_{part}$ will play an important role in the quantization
of the system (see Sect. 7).

In section 3 we showed that the residual symmetry 
 contain the translations local in time.
 The conserved quantities $P\sp{i}$, which vanish according to 
 (4.23) and (4.29) in the physical phase space are the generators
of this symmetry in an extended phase space. 
 The same result has been obtained
  for $(2+1)$-dimensional gravity  [35] and
 by one of the present authors (PSC) in the one--dimensional case [2], [3]. 
For the similar case of local (in time) rotations 
 it has been shown that the corresponding rotation
 generator is a first class constraint
 which  vanish in the physical part  of the phase space 
   (cp. [36], [37]).

\section{Classical Particle Dynamics (N Body Problem)}

\renewcommand{\theequation}{\thesection.\arabic{equation}}

\subsection{General Properties of the Solutions}

Let us look at the classical equations of motion and their solutions.
The equations to solve are then
\begin{equation}
\xi_{\alpha}\sp{\underline{a}}\, =\, E_{\alpha,i}\sp{\underline{a}}
 \dot x\sp{i,\underline{a}}\,+\,
 E_{\alpha,0}\sp{\underline{a}}
\end{equation}
and
\begin{equation}
\dot \xi_{\alpha}\sp{\underline{a}}\cdot E_{\alpha,i}\sp{\underline{a}}
\,+\,\xi_{\alpha}\sp{\underline{a}}\cdot F_{\mu i,\alpha}\sp{\underline{a}}
\dot x\sp{\mu,\alpha}\,=\,0
\label{two}
\end{equation}
for the particle coordinates $\{\vec{x}\sp{\alpha}(t)\}$ 
after the insertion of the solution (4.2) 
for $E_{\alpha,\mu}\sp{\underline{a}}$ into these equations.

However, from (4.1) we obtain for $ F_{\mu \nu}\sp{\underline{a}}$
\begin{equation}
F_{\mu \nu}\sp{\underline{a}}\,=\,-{\lambda\over 2}\,\epsilon_{\mu\nu\rho}
\sum\limits_{\beta}\xi_{\beta}\sp{\underline{a}}\,\dot x\sp{\rho,\beta}
\,\delta(\vec{x}-\vec{x}\sp{
\beta}).
\end{equation}
Thus the second term in (5.2), due to the antisymmetry of the 
$\epsilon$ tensor, can be rewritten as

\begin{equation}
\xi_{\alpha}\sp{\underline{a}}\cdot F_{\mu i,\alpha}\sp{\underline{a}}
\dot x\sp{\mu,\alpha}\,=\,
-{\lambda\over 2}\epsilon_{\mu i\rho}\sum\limits_{\beta\ne \alpha}
\xi_{\beta}\sp{\underline{a}}\, \
\xi_{\alpha}\sp{\underline{a}}\,
\dot x\sp{\rho,\beta}\dot x\sp{\mu,\alpha}\delta(\vec{x}_{\alpha}-
\vec{x}_{\beta})
\end{equation}
which is infinite for coinciding particle positions and vanishes otherwise. 
Therefore our configuration space contains only 
noncoinciding 
particle positions.

From eq. (5.2) and (5.4) we have
\begin{equation}
\dot \xi_{\alpha}\sp{\underline{a}}\cdot E_{\alpha,i}\sp{\underline{a}}\,=\,0,
\label{NNew}
\end{equation}
which leads, for points in the configuration space where the metric 
is non-degenerate, to
\begin{equation}
\dot \xi_{\alpha}\sp{\underline{a}}\,=\,0.
\label{New}
\end{equation}

Using this result we obtain for $\xi_{\alpha}\sp{\underline{a}}$ 
given by (1.3) from 
(4.2) and (4.10--12)
\begin{equation}
\xi_{\alpha}\sp{\underline{a}}\,=\,\dot x_{\alpha}\sp{\underline{a}}
-v\sp{\underline{a}}-{\lambda\over 4\pi}\sum\limits_{\beta\ne \alpha}\,
\xi_{\beta}\sp{\underline{a}}\left(\dot x_{\alpha\beta}\sp{\underline{b}}
\cdot \partial_b \Phi(\vec{x}\,_{\alpha\beta})\right),
\label{new2}
\end{equation}
where we have defined
\begin{equation}
x_{\alpha\beta}\sp{\underline{b}}:= x_{\alpha}\sp{\underline{b}}
- x_{\beta}\sp{\underline{b}}.
\end{equation}
For the particular case of a particle motion on a line we find, 
from the last two
expressions, that
\begin{equation}
\ddot x_{\alpha}\sp{\underline{a}}\,=\,0
\end{equation}
{\it i.e.} the free motion.

This is also easily seen from (see (1.3c) in [26])
\begin{equation}
\dot x_{\alpha\beta}\sp{\underline{a}}\cdot \partial_a \Phi(x_{\alpha\beta})\,
=\, {\vec{x}}\,_{\alpha\beta}\wedge {\dot{\vec{x}}\, 
_{\alpha\beta}
\over |      \vec{x}_{\alpha\beta}|\sp2}\,=\,0
\end{equation}
as $ \vec{x}_{\alpha\beta}$ 
 and $\dot{\vec{x}}_{\alpha\beta}$ are parallel for
  a motion along a line.

We should say also a few words about the 
degenerate case {\it i.e.} when (5.6) is not 
  necessarily satisfied.

Then we have for $\xi_{\alpha}\sp{\underline{a}}$ from 
(4.2) and (4.10--12)

\begin{equation}
\xi_{\alpha}\sp{\underline{a}}\,
=\dot x_{\alpha}\sp{\underline{a}}\,-\,v\sp{\underline{a}}\,-\,
{\lambda\over 4\pi}\sum_{\beta\ne\alpha}\,\xi_{\beta}\sp{\underline{a}}
\left( \dot x_{\alpha\beta}\sp{\underline{b}}\cdot
 \partial_b\Phi(\vec{x}_{\alpha\beta}
)\right)\,
-\,{\lambda\over 4\pi}\sum_{\beta\ne\alpha}\,\dot\xi_{\beta}\sp{\underline{a}}
\left( \Phi(\vec{x}_{\alpha\beta})-\Phi(0)\right),
\end{equation}
where $\Phi(0)$ has to be defined by an appropriate regularization procedure.
But for any such procedure the last term in the last formula breaks 
rotational invariance. Thus, although $\tilde E_0\sp{\underline{a}}(\vec{x},t)$
is rotationally invariant this is not the case for
\begin{equation}
\tilde E_{\alpha,0}\sp{\underline{a}}\,=\,\lim_{\vec{x}\rightarrow 
\vec{x}_{\alpha}} \tilde{E}\,_{0}\sp{\underline{a}}(\vec{x},t)
\end{equation}
if (5.6) does not hold.

Hence the degenerate case can be considered as
  unphysical and further will not be discussed.

\subsection{Reduced Hamiltonian/Lagrangian for the 
Non--Degenerate Case}
Applying the Legendre transformation to the total Lagrangian
 given by (4.21)
we find the following Hamiltonian 
\begin{eqnarray}
H\,=\,&-\int \,dx\sp2\,
\tilde{E}\,_0\sp{\underline{a}}\cdot \left( {2\over \lambda}
\epsilon\sp{ij}\partial_i\,
\tilde{E}\,_j\sp{\underline{a}}\,+\,+
\sum\limits_{\alpha}\xi_{\alpha}\sp{\underline{a}}\,
\delta(\vec{x}-\vec{x}_{\alpha})\right)\nonumber\\
&+\,{1\over 2}\sum\limits_{\alpha}\xi_{\alpha}\sp{\underline{a}}
 \xi_{\alpha}\sp{\underline{a}}\,+\,v\sp{\underline{a}}\cdot \sum\limits_{\alpha}
\xi_{\alpha}\sp{\underline{a}}.
\end{eqnarray}

The gauge field $ \tilde{E}\,_0\sp{\underline{a}}$ 
 is a Lagrange multiplier field, whose
 variation gives the constraint
\begin{equation}
\epsilon\sp{ij}\partial_i\,\tilde{E}\,_j\sp{\underline{a}}
(\vec{x},t)\,= \,-{\lambda\over 2}
\sum\limits_{\alpha}\xi_{\alpha}\sp{\underline{a}}\,\delta(
\vec{x}-\vec{x}_{\alpha}(t))
\end{equation}
which corresponds to the $\mu=0$ term in (4.1).

When if we put this expression into the Hamiltonian we find
\begin{equation}
H^{(N)}\,=\,{1\over2} \sum\limits_{\alpha}\xi_{\alpha}\sp{\underline{a}}
 \xi_{\alpha}\sp{\underline
{a}}\,+\,v\sp{\underline{a}}\cdot \sum\limits_{\alpha}
\xi_{\alpha}\sp{\underline{a}}.
\label{new1}
\end{equation}

This is a free Hamiltonian with a constraint and the 
 dynamics  is contained in the non-trivial
symplectic structure.
\footnote{Compare with [34] where
 a similar discussion of the Chern--Simons electrodynamics for
   point particles is given.}

In order to complete the Legendre transformation we have to express the 
auxiliary variable $\xi_{\alpha}\sp{\underline{a}}$ in terms of the canonical
variables $\{x_{\alpha}\sp{\underline{a}},\,p_{\alpha}\sp{\underline{a}}\}$.
This can be done in a unique way iff
\begin{equation}
det\left({\partial p_{\alpha}\sp{i}\over \partial \xi_{\beta}\sp{j}}\right)
\ne0,\end{equation}
where
\begin{eqnarray}
&p_{\alpha}\sp{{i}}:=\,{\partial L\over \partial 
\dot x_{\alpha}\sp{{i}}}\,=\, 
\xi_{\alpha}\sp{\underline{a}}
\cdot E_{\alpha \, i}\sp{\underline{a}}\,=\nonumber\\
=&\xi_{\alpha}\sp{i}\,-\,{\lambda\over 4\pi}
\sum\limits_{\beta\ne\alpha}(\xi_{\alpha}\sp{\underline{a}}\cdot
\xi_{\beta}\sp{\underline{a}})\partial_i 
\Phi(\vec{x}\,_{\alpha\beta}).\label{defp}
\end{eqnarray}

The condition (5.6) defining the non-degenerate metric now 
arises as a consequence of the EOM. When
 we apply an inverse Legendre
transformation to (5.15) we get

\begin{equation}
L_{red}\,=\,\sum_{\alpha}p_{\alpha}\sp{\underline{a}}
 \cdot \vec{x}\,_{\alpha}
\sp{\underline{a}}\,-\,H_{red}\, .
\end{equation}

Varying $S_{red}$ with respect to $x_{\alpha}\sp{{i}}$
we obtain using (5.17)
\begin{equation}
\sum_{\beta}{\partial p_{\alpha}
\sp{{i}}\over \partial \xi_{\beta}
\sp{\underline{a}}}\,\dot \xi_{\beta}\sp{\underline{a}}\,=\,0
\end{equation}
which leads only to the trivial solution
\begin{equation}
\dot\xi_{\alpha}\sp{\underline{a}}\,=\,0,
\label{new3}
\end{equation}
if (5.16) holds. We conclude therefore that the existence of the Legendre
transformation implies a non-degenerate metric.

Varying $S_{red}$ with respect to $v\sp{\underline{a}}$
 and 
$\xi_{\alpha}\sp{\underline{a}}$ we obtain, respectively, the constraints
\begin{equation}
\sum\limits_{\alpha}\xi_{\alpha}\sp{\underline{a}}\,=\,0,
\end{equation}
in accordance with the relation (4.23) obtained in geometric way
 by
 suitable gauge fixing
and
\begin{equation}
\xi_{\alpha}\sp{\underline{a}}\,=\,\dot x_{\alpha}\sp{\underline{a}}
\,
-\,v\sp{\underline{a}}\,-\,{\lambda\over 4\pi}
\sum\limits_{\beta \ne\alpha}(\xi_{\beta}\sp{\underline{a}}\cdot
\left(\dot x_{\alpha\beta}\sp{\underline{b}}
\cdot\partial_b \Phi(\vec{x}\,_{\alpha\beta})
\right).
\label{last}
\end{equation}
Alternatively, we may consider the Hamiltonian (5.15)
in the $\{x_{\alpha}\sp{\underline{a}},\,\xi_{\alpha}
\sp{\underline{a}}\}$-space.
In such a case the free Hamiltonian 
 (5.15)    will be endowed with 
the nontrivial symplectic structure.

\subsection{Transformation to flat coordinates}

Let us recall from Sect. 2  
 that the space parts of our dreibeins 
  $E_i\sp{\underline{a}}$ determine the metric $g_
{ij}$ in the two-dimensional space
\begin{equation}
g_{ij}(\vec{x},t):=\,(E_i\sp{\underline{a}}(\vec{x},t)
\cdot E_j\sp{\underline{a}}(\vec{x},t))
\end{equation}
with the line element given by
\begin{equation}
ds\sp2\,=\,g_{ij}\,dx\sp{i}\,dx\sp{j}.
\end{equation}

But according to (4.2) the $ E_i\sp{\underline{a}}$ are singular
due to the formula (we  put
$\tilde{\Lambda}^{\underline{a}}\equiv y^{\underline{a}}$)
\begin{equation}
E_i\sp{\underline{a}}\,=\,\partial_i\,y\sp{\underline{a}}(\vec{x},t)
\end{equation}
with the singularities located at the  particle positions
$\{x_{\alpha}\sp{\underline{a}}(t)\}, (\alpha = 1, \ldots N)$.
  We see that our metric 
is locally flat\footnote{Compare with (2+1) dimensional gravity
(cp. [22])} and we have
\begin{equation}
ds\sp2\,=\,(dy\sp{\underline{a}}\cdot dy\sp{\underline{a}})
\end{equation}
in $R\sp2-\{x_{\alpha}\sp{\underline{a}}(t)\},  (\alpha = 1, \ldots N)$.

We can now show that
 the variables 
 $\{y_{\alpha}\sp{\underline{a}}(t)\}$ are canonically 
conjugate to the covariantized velocities 
 $\{\xi_{\alpha}\sp{\underline{a}}(t)\}$. To show this we
consider $F(\{\xi_{\alpha}\sp{\underline{a}},x_{\alpha}\sp{\underline{a}}\})$
defined as follows:
\begin{equation}
F:=\,\sum\limits_{\alpha}
\xi_{\alpha}\sp{\underline{a}}
\cdot x_{\alpha}\sp{\underline{a}}\,-\,{\lambda\over 8\pi}
\,\sum\limits_{\alpha,\beta\atop \alpha \ne \beta}
 (\xi_{\beta}\sp{\underline{a}}\cdot
\xi_{\alpha}\sp{\underline{a}})\,
\phi(\vec{x}\,_{\alpha\beta})
\end{equation}
 $F$ is a generating function for the canonical
transformation
\begin{equation}
\{ x_{\alpha}\sp{\underline{a}}, 
p_{\alpha}\sp{\underline{a}}\}\quad \rightarrow
\quad \{ y_{\alpha}\sp{\underline{a}}, \xi_{\alpha}\sp{\underline{a}}\}
\end{equation}
at the points in phase space for which
\begin{equation}
det\left({\partial \sp2 F\over \partial \xi_i\partial x_j}\right)\,\ne\,0.
\label{new4}
\end{equation}
Indeed,  the relation (5.17) takes the form
\begin{equation}
p_{\alpha, i} = {\partial  F\over \partial  x_{\alpha, i} }
\label{new53a}
\end{equation}
and from (5.25), (4.2) and (4.8--11) 
it follows
\begin{equation}
y_{\alpha, i} = {\partial  F\over \partial  \xi_{\alpha, i} }
\label{new53aa}
\end{equation}

Thus we conclude that the motion is free
in the part of the phase space $ \{ y_{\alpha}\sp{\underline{a}}, 
\xi_{\alpha}\sp{\underline{a}}\}_{\sum\limits_{\alpha}
\xi_{\alpha}\sp{\underline{a}}=0}$ 
 restricted by  the condition (5.29) 
 and with  substracted  coinciding particle
positions.

\section{Two Body Problem}

\subsection{Symplectic Structure}


We define:

\begin{equation}
\xi\sp{\underline{a}}:= {1\over 2} \left(
\xi_1\sp{\underline{a}}\, -\,\xi_2\sp{\underline{a}}\right) \, ,
\qquad 
x\sp{\underline{a}}:=\, x_1\sp{\underline{a}}\,
-\,x_2\sp{\underline{a}}\, ,
\qquad 
p\sp{\underline{a}}:=\,{1\over2}( p_1\sp{\underline{a}}
\, -\,p_2\sp{\underline{a}})\, .
\end{equation}
Then using the constraint (5.21) we get from (5.15)
\begin{equation}
H^{(2)}\,=\, \xi\sp{\underline{a}}\cdot \xi\sp{\underline{a}}
\label{2.29}
\end{equation}
and
\begin{equation}
p_i\,=\, 
\xi_i\,+\,{\lambda\over 4\pi}
(\xi\sp{\underline{a}}\cdot \xi\sp{\underline{a}})
\partial_i\,\Phi(\vec{x})
\, .
\label{eqxx}
\end{equation}

 The
Hamiltonian equations take the form

\setcounter{equation}{0}
\renewcommand{\theequation}{\thesection.4\alph{equation}}
\begin{equation}
\dot{x}\sp{i}\,=\, {\partial H\over \partial p\sp{i}}\,=\,{2}
\left(\xi\sp{\underline{a}}\cdot {\partial \xi\sp{\underline{a}}\over 
\partial p\sp{i}}\right),
\end{equation}

\begin{equation}
\dot{p}_{i}\,=\,- \,  {\partial H\over \partial x_{i}}\,= - \,{2}
\left(\xi\sp{{a}}\cdot {\partial \xi\sp{{a}}\over 
\partial x\sp{i}}\right).
\end{equation}

Using 
  (\ref{eqxx})  we have
\renewcommand{\theequation}{\thesection.5\alph{equation}}
\setcounter{equation}{0}
\begin{equation}
\xi\sp{\underline{a}}\cdot {\partial \xi\sp{\underline{a}}\over 
\partial p\sp{i}} 
\,=\,{\xi_i\over 
1+ \displaystyle{\lambda\over 2\pi}
(\xi\sp{\underline{a}}\cdot \partial_a\Phi)},
\end{equation}

\begin{equation}
\xi\sp{{a}}\cdot {\partial \xi\sp{{a}}\over 
\partial x\sp{j}}
\,=\,\ \, 
-\, {\lambda \over 4\pi} \
{\left(\xi_a  \xi^a \right) \xi^i\partial_i \partial_j \Phi \over 
1+  \displaystyle{\lambda\over 2\pi}(\xi\sp{{a}}\cdot 
\partial_a\Phi)}.
\end{equation}
\renewcommand{\theequation}{\thesection.\arabic{equation}}
\setcounter{equation}{5}

Taking the time derivative of 
 (\ref{eqxx})  and using (6.4--5) we get
\begin{equation}
\dot{\xi}_{i} + {\lambda \over 2\pi} \partial_{i}
\Phi(x) 
\xi_{j}\dot{\xi}_{j} = 0 \, .
\end{equation}

Let us now illustrate for $N=2$ the procedure leading to the
free motion (5.20).
Instead of the canonical variables ($x_i, p_i$) we can
use the variables ($x_i, \xi_i$). Then the Lagrangian 
obtained from the Hamiltonian 
  (6.2)  would have had the form:
\begin{eqnarray}
L & = &  p_l (\xi_i , x_i )\cdot \dot{x}_l - H 
\cr
& = & \left( \xi_l + {\lambda \over 4\pi} \xi^2
\partial_l \Phi(x)\right)\dot{x}_{l} -  \xi^2\, .
\end{eqnarray}
The variation with respect to $\xi_i$ is given by the expression
\begin{equation}
\xi_i = \displaystyle { { 1\over 2} \dot{x}_{i} \over 1 
- \displaystyle{\lambda\over 4\pi} \left(
\dot{x}_{j}\partial_j \Phi \right) },
\end{equation}
which is equivalent to the Hamiltonian eq. (6.4a) with the
insertion of (6.5a). Inserting (6.8) in (6.7) we get
\begin{equation}
L = {1\over 4} \  \displaystyle{\dot{x}^{2}_{l} \over
  \displaystyle 
  1 - \displaystyle{\lambda\over 4\pi} \left( \dot{x}_{l} \partial_l 
\Phi
\right) } \, .
\end{equation}
In particular if we observe that
\begin{eqnarray}
\det\left( {\partial^2 L \over \partial \dot{x}_i \partial \dot{x}_j 
}\right)
&= &
{1\over 4} \left(1 - {\lambda\over 4\pi} \left( \dot{x}_{l} 
\partial_l \Phi
\right) \right)^{-4}
\cr
&= &
{1\over 4} \left(1 + {\lambda\over 2\pi} 
 \xi_{l} \partial_l \Phi
 \right)^{+4}\, ,
 \end{eqnarray}
we see that if the 
 velocities are expressible in terms of the canonical variables
then
 from (6.6) it follows that
\begin{equation}
\dot{\xi}_l = 0\, .
\end{equation}
Using the Hamiltonian 
 (6.2)  we get for the pair of
noncanonical variables the Hamilton equations (6.4a) in the form
$\dot{x}_l = \{ x_l , H \}$ as well as (6.11) given by 
$\dot{\xi}_l = \{ \xi_l , H \}$, provided that we assume the following
nonstandard symplectic structure:

\begin{equation}
\left\{ x^i , x^j \right\} = \left\{ \xi_i, \xi_j \right\} = 0,
\end{equation}

\begin{equation}
  \left\{ x^i , \xi_j  \right\}= \delta^{i}_{\ j}
- {{\lambda\over 2\pi} \xi^i \partial_j \Phi \over
1 +  \displaystyle
{\lambda \over 2\pi } \left( \xi^a \partial_a \Phi \right) }.
\end{equation} 

It is easy to check  that the Poisson brackets (6.12-6.13) satisfy
the Jacobi identity.

\subsection{Two Conserved Angular Momenta}

For our system we have two conserved scalar angular momenta (cp. section 4.2):

i) If we define (in $d=2$ $\phantom{aa}$  $\vec{a} \wedge \vec{b} = \epsilon_{ij}a^i
b^j$  is a scalar)
\begin{equation}
l:= \,\vec{x}\,\wedge {\vec{p}}\,
\,=\,
\vec{x}\,\wedge \vec{\xi}\,
+{\lambda\over 4\pi} H
\label{ang}
\end{equation}
we find that  $l$ is conserved 

ii) Second conserved angular momenta $\overline{l}$ is the following

 \begin{equation}
 \bar l:=\vec{x}\,\wedge \vec{\xi}\,
 \end{equation}
 because 
\begin{equation}
{d\over dt}\bar l\,=\,\dot{\vec{x}}\,\wedge 
\vec{\xi}\,
\,=\,0.
\end{equation}

Using the relation
\begin{equation}
\xi\sp{\underline{a}}\cdot \partial_a\,\Phi\,
=\,- \vec{\xi}\,\wedge
\vec{\nabla}\,\ln r\,=\,{\bar l\over r\sp2},
\end{equation}
where $r:=|\vec{x}|$,  we see that (6.8) can be 
rewritten as:
\begin{equation}
\dot x\sp{\underline{a}}\,=\,{ 2\xi\sp{\underline{a}}\over
1+{\lambda\bar l\over 2\pi r\sp2}}.
\end{equation}

\renewcommand{\theequation}{\thesection.\arabic{equation}}
\setcounter{equation}{18}
Note that if $\vec{x}(0)$ is parallel to 
$\vec{\xi}$
we have $\bar l=0$ and, in consequence, a free motion on a line.

 For $\lambda \overline{l} < 0$ it is convenient to
introduce the quantity 
 \begin{equation}
 r^{2}_{0} : =  { | \lambda \overline{l} |  \over 2 \pi }
 \end{equation}
We see from (6.18) that the relative 
two--body
 problem separates in this case into a motion within the interior
region given by
\renewcommand{\theequation}{6.20\alph{equation}}
\setcounter{equation}{0}
\begin{equation}
r < r _0
\end{equation}
and a motion in the exterior region given by

\begin{equation}
r > r _0.
\end{equation}

\renewcommand{\theequation}{\thesection.\arabic{equation}}
\setcounter{equation}{20}
\subsection{Structure of the classical phase-space $M$}
Let us denote by $M$  the phase space for the canonical variables
($\vec{x}, \vec{p}$). First of all, let us observe that
$$ M\ne R\sp{2}\otimes R\sp2.$$

To show this we  start with (6.3) rewritten as
\begin{equation}
\vec{\xi}\,=\,\vec{p}\,-\,{\lambda\over 4\pi}\,
H\,\vec{\bigtriangledown} \phi.
\end{equation}

Squaring it we get
\begin{equation}
\left({\lambda\over 4\pi}{H\over r}-\left({2\pi r\over \lambda}+
{l\over r}\right)\right)\sp2+
\vec{p}\sp2-\left({2\pi r\over \lambda}+
{l\over r}\right)\sp2 = 0,
\label{eq4}
\end{equation}
where we have used the definition (6.14) of $l$.
Thus we must have
\begin{equation}
\vec{p}\sp2-\left({2\pi r\over \lambda}+
{l\over r}\right)\sp2\,\le\,0
\end{equation}
and so
(with $l=rp\sin{\varphi}$)
\begin{eqnarray}
r\,\ge\, &{\lambda\over 2\pi}\,p\,(1-\sin{\varphi}), 
\qquad\mbox{for}\quad \lambda>0, \nonumber\\
r\,\ge\, &{\lambda\over 2\pi}\,p\,(1+\sin{\varphi})
\qquad\mbox{for}\quad \lambda<0,
\label{eq5}
\end{eqnarray}
where the equality sign holds for $r=r_0$.
Therefore the map $(\vec{x},\vec{\xi})\rightarrow (\vec{x},\vec{p})$
with $(\vec{x},\vec{\xi})\in R\sp2\otimes R\sp2$ leads to
$(\vec{x}, \vec{p})\in M\ne R\sp2\otimes R\sp2$.

From (6.22) we may look for the Hamiltonian $H$ as a function of
the canonical variables $(\vec{x},\vec{p})$
\begin{equation}
H = {4\pi r\over \lambda} \left[ 
{2\pi r \over \lambda} + {l\over r} \pm
\sqrt{ \left( 
{2\pi r \over \lambda} + {l \over r} \right)^{2}
- \vec{p}^{\ 2}} \,  \right]
\end{equation}
with the $- (+)$ sign in front of the second term in (6.25) for
$\left( 
{2\pi r \over \lambda} + {\bar{l} \over r} \right) > 0$
 $(< 0)$.
 Let us note, however, that due to  formula (6.14), the
RHS of (6.25) depends through $l$ also on the energy 
${H}$. Writing (6.22)  as 
\begin{equation}
\left( 
{\bar{l}\over r}  + \,  {2\pi r \over \lambda}  \right)^{2}
 + \vec{p}^{\ 2} -
 \left( 
{2\pi r \over \lambda} + {\overline{l} \over r} 
+ {\lambda \over 4\pi r} H \right)^{2} = 0
\end{equation}
we get
\begin{equation}
H = {4\pi r     \over \lambda}
\left[ - 
\left( 
{2\pi r \over \lambda} + {\overline{l} \over r} \right)
\pm 
\sqrt{\vec{p}^{\ 2} +
\left( 
{2\pi r \over \lambda} + {\overline{l} \over r} \right)^{2} 
}\ \right].
\end{equation}
The formulae (6.25) and (6.27) can be used alternatively in the
quantization procedure,  depending on the choice of the
eigenstates of angular momenta ($l$ in the case of eq. (6.25)
and $\overline{l}$ in the case of eq. (6.27)).

\subsection{Classical Trajectory}
To find the classical trajectories we can make the ansatz
\begin{equation}
\vec{x}(t)\,=\,\alpha(t) \, \vec{\xi}\,+\,\beta(t)\, \vec{\xi}_\perp,
\label{decompa}
\end{equation}
where we have written $\vec{x}$ for $x\sp{\underline{a}}$
and similarly to $\vec{\xi}$ and defined
$\vec{\xi}_{\perp}$ as a unit vector perpendicular to $\vec{\xi}$.
Then 
\begin{equation}
\vec{\xi}\wedge\vec{\xi}_\perp\,=\,|\vec{\xi}|\,=\,\xi.
\end{equation}
As $\xi$ is conserved, the conservation of $\bar l$ tells us
that $\beta$ is constant
and so (\ref{decompa}) becomes
\begin{equation}
\vec{x}(t)\,=\,\alpha(t)\, \vec{\xi}\,+\,\beta\,\vec{\xi}_\perp
\end{equation} 
with a constant $\beta$. So the problem has been reduced to 
 finding $\alpha(t)$.
To do this we choose a frame in which $\vec\xi=\xi\,\vec{e}_x$. Then
$\vec{\xi}_\perp=\vec{e}_y$ and we see that
\begin{equation}
\phi(\vec{x})\,=\,\arctan\left({\beta\over \alpha(t)\xi}\right).
\end{equation}

Next, as in the previous section, we introduce $\vec{y}$ 
\begin{equation}
\vec{y}\,=\,\vec{x}\,+\,{\lambda\over 2\pi}\,\vec{\xi}\,\phi(\vec{x}),
\end{equation}
which is canonically conjugate to $\vec{\xi}$, and find that 
using the last two formulae we see that
$\alpha(t)$ satisfies
\begin{equation}
2\, \vec{\xi}t\,+\,\vec{y}(0)\,=\,\alpha(t)\vec{\xi}\,+\,\beta\vec{\xi}_\perp
\,+\,{\lambda\over 2\pi}\,
  \vec{\xi} \, \arctan\left({\beta\over \alpha(t)\xi}\right),
\end{equation}
{\it i.e.}
we arive at a fixed point equation for $\alpha(t)$
\begin{equation}
\alpha(t)\,=\,-{\lambda\over 2\pi}\,\arctan\,
\left({\beta\over \alpha(t)\xi}\right)
\,+\, 2 t\,+\,c,
\label{sol}
\end{equation}
with $c$ determined by the initial conditions.

To discuss the conditions on the existence of $\alpha(t)$ which solve
(\ref{sol}) we rescale it by introducing
\begin{equation}
\tau= 2 \, {t\xi\over \beta},\quad 
 k=-{\lambda \xi\over 2\pi \beta},\quad \mbox{and}\quad
 g(\tau)= 2\, {\alpha(t)\xi\over \beta}.
\end{equation}
Then after a time translation $t+c\rightarrow t$ our equation (\ref{sol})
becomes
\begin{equation}
g\,=\,f(g)\,+\,\tau,
\label{Eq}
\end{equation} 
where
\begin{equation}
f(g)\,=\,k\arctan\left({1\over g}\right).
\end{equation}

To discuss the solvability of (6.36) it is convenient 
to redefine $f$ to be given by
\begin{equation}
f(g)\,=\,-k\arctan {g}
\end{equation}
and perform a further time translation so that the equation for $g$ 
is still of the form (6.36). Then putting
\begin{equation}
g\,=\,\nu\,+\,\tau
\end{equation}
we find that (6.36) becomes
\begin{equation}
\nu\,=\,-k\int_0\sp{\nu+\tau}\,dx\,{1\over 1+x\sp2}\,=:\,g_{\tau}(\nu).
\end{equation}

Note that this implies that
\begin{equation}
\nu(-\infty)\,=\,k\,{\pi\over 2}
\end{equation}
and
\begin{equation}
\nu(\infty)\,=\,-k\,{\pi\over 2}
\end{equation}
independently of the value of $k$.

Note further that
\begin{equation}
{\partial g_{\tau}\over \partial \nu}(\nu)\,=\,-k\,{1\over 1+(\nu+\tau)\sp2}
\end{equation}

Let us consider first the case of $k>-1$.

We rewrite (6.40) as
\begin{equation}
G_{\tau}(\nu):=\,\,\nu\,-\,g_{\tau}\,=\,0.
\label{cc}
\end{equation}

Then from (6.40) we have
\begin{equation}
G_{\tau}(\pm \infty)\,=\,\pm \infty.
\label{bb}
\end{equation}

Differentiating (6.26) with respect to $\tau$ we find
\begin{equation}
\dot \nu(\tau)\,=\,-{k\over (\nu(\tau)+\tau)\sp2+1+k}
\end{equation}

From (6.43) we conclude that $G_{\tau}(\nu)$ is 
a monotonically increasing function of $\nu$. 
 Given (6.45) and (6.46) this 
implies that there is a unique and smooth 
 solution $\nu(\tau)$ of (6.44)
 for each $\tau\in R\sp1$.

For $ k < -1$ we have a problem as 
$\nu$ develops a discontinuity.  
This can be seen by plotting the various terms in (\ref{Eq}).
Thus we see that for $k>-1$ we have a well defined 
$g(\tau)$ while for $ k <-1$ the function $g(\tau)$
``jumps" at some value of $\tau$ showing that the $x$ space is not the physical
configuration space in this case. 

Let us discuss this in more detail.

 Eq. (6.46)  shows
 the smoothness of $\nu(\tau)$ for all values of $k>-1$. For 
$k<-1$, however, $\nu(\tau)$ has an infinite slope at $\tau=\pm\tau_0$, where
\begin{equation}
\tau_0:=\,-\sqrt{\vert k\vert -1}\,+\,\vert k\vert
 \int_0\sp{\sqrt{\vert k\vert -1}}
\,dx\,{1\over 1+x\sp2}.\end{equation}

But due to the symmetry $\nu(-\tau)=-\nu(\tau)$ it is 
 sufficient to consider the case of $\tau=\tau
_0$.
To do this we note that if we start at $\tau=-\infty$ we have a smooth function
$\nu(\tau)$ for $\tau<\tau_0$. Close to $\tau=\tau_0$  we have
\begin{equation}
\lim_{\epsilon\rightarrow +0}\,\dot \nu(\tau_0-\epsilon)\,=\,+\infty
\end{equation}
with
\begin{equation}
\nu(\tau_0-\epsilon)\,=\,-\tau_0\,-\,\sqrt{\vert k\vert-1}\,+\,\epsilon
\end{equation}
thus showing that $\nu(\tau)$ jumps at $\tau=\tau_0$ to
\begin{equation}
\nu(\tau_0+\epsilon)\,>\,\nu(\tau_0-\epsilon).
\end{equation}

However, the states $\nu(\tau_0\pm \epsilon)$ are physically equivalent;
{\it i.e.} they have to be identified in the configuration space $C_{phys}$. 
(6.33) shows that $C_{phys}$ is determined by $y$-space defined by (6.32).
To see this we recall from section 5.3 that
  the map $\vec{x}\rightarrow \vec{y}$ is part of a canonical
transformation iff
\begin{equation}
\det\left({\partial\sp2 F\over \partial x_i\partial \xi_j}\right)\,\ne\,0
\end{equation}
which for $N=2$ is equivalent to
\begin{equation}
1\,+\,{\lambda\over 2\pi}\,\xi\sp{i}\,\partial_i \phi\,\ne\,0
\label{aa}
\end{equation}
and so
\begin{equation}
(\nu\,+\,\tau)\sp2\,+\,1\,+\,k\,\ne\,0.
\end{equation} 

But for the whole trajectory $\nu(\tau)$ we have
(with the identification $\nu(\tau_0-\epsilon)\sim \nu(\tau_0+\epsilon)$)
\begin{equation}
(\nu\,+\,\tau)\sp2\,+\,1\,+\,k\,>\,0,
\end{equation} 
which, due to (6.17)  is equivalent to
\begin{equation}
r\sp2> r_0\sp2   
 \qquad \mbox{for}\quad \lambda\bar{l}<0.
\end{equation}

Thus we have a one-to-one correspondence between the $y$-space and $C_{phys}$
for $r>r_0$. 

Within the interior of the region $r<r_0$ we obtain from (6.40) a solution 
valid only for a finite time interval. Such a situation is well
known in General Relativity 
(cp. [38]) and the literature cited there).

From the physical point of view the case
$k<-1$ is the most interesting as then the configuration space consists 
of two non-communicationg regions; 
the exterior ($r>r_0)$ and the interior ($r<r_0)$. Thus, in the 
next section we will concentrate on the quantization
of this case.

\section{Quantization of the Two--Body Problem}
\subsection{Nonstandard Schr\"{o}dinger Equation}

The relation
(6.3), 
 after the substitution of  
 (6.2),
takes the form ($H\equiv H^{(2)}$):

\begin{equation}
\vec{\xi}\,=\,\vec{p}\,-\,{\lambda\over 4\pi} \, H
\,\vec{\bigtriangledown} \, \Phi.
\end{equation}

\setcounter{equation}{0}
\renewcommand{\theequation}{\thesection.2\alph{equation}}

Squaring it and using again (6.2) we obtain
\begin{equation}
H = \vec{p}\, ^{2} - {l^2 \over r^2 } +
 { \overline{l}\, ^2 \over r^2},
\label{eq4}
\end{equation}
where we have used the definition  (6.14) of $\overline{l}$.   
 
Let us note  that
 \begin{equation}
 \overline{l} = l - {\lambda \over 4 \pi } H
 \end{equation}
 {\it i.e.} (7.2a) gives us a {\bf quadratic} equation for $H$ (cp. section 6.3).

\setcounter{equation}{2}
\renewcommand{\theequation}{\thesection.\arabic{equation}}

We quantize the problem by considering a Schr\"odinger-like equation 
\begin{equation}
i\hbar {\partial \psi(\vec{x},t) \over 
\partial t} =    \hat{H} \, \psi (\vec{x},t) =
\left[
\hat{\vec{p}}\, \sp2- {l^2  \over r^2} +
{1\over  r^2 } \, \overline{l}\, ^{2} \right]
\Psi(\vec{x},t)
\label{eq6}
\end{equation}
in which the operators $\hat H$ and $\hat{\vec{p}}$ 
  are defined  by  the usual
quantization rules
\begin{equation}
\hat H\,= \, i\hbar {\partial\over \partial t},\qquad
 \hat{p}_{l}\,= \,
{\hbar\over i}\partial_l \, .
\end{equation}

We see that the equation (7.3) describes a nonstandard form of 
a time dependent  Schr\"{o}dinger equation, with its right hand side
  containing  both the first and second time derivatives (entering 
through $\overline{l}$).

For the stationary case, {\it i.e.}
 when $\Psi(\vec{x},t)=\Psi_E(\vec{x})e\sp{{iEt\over \hbar}}$
we can use the angular-momentum basis and put
\begin{equation}
\Psi_{E,m}\,=\,f_{E,m}(r ) \, e\sp{im\varphi}
\end{equation}
where $m$ is an integer, and find that $f_{E,m}$ satisfies a nonstandard
 time independent  Schr\"odinger equation
\begin{equation}
\left[-\hbar\sp2\left(\partial_r\sp2 
+ {1\over r}\partial_r 
-{\bar{m}\sp2\over r\sp2}\right)-E\right]f_{E,m}(r) =0,
\label{sch}
\end{equation}
where in consistency  with (7.2b)  we have defined
\begin{equation}
\hbar \bar{m} :\,=\,\hbar m-{\lambda\over 4\pi} E
\label{bar}
\end{equation}
{\it i.e.} $\hbar \bar{m}$ is an eigenvalue of ${\overline{l}}$.

A characteristic feature of the Schr\"odinger equation (7.6) is the 
appearance
of the noncanonical angular momentum $\hbar\bar{m}$ with $\bar{m}$ not 
being an integer.  
{\it i.e.} our two-body system carries a fractional orbital angular
momentum (see the discussion in [27] or [39]). It has been shown
in section 4.2 
that $J= {\overline{l}}$ is equal to the total
angular momentum of the particle + field system with a
nonvanishing angular momentum of the fields. We note that a form of (7.7)
shows a great similarity bteween our two-particle state and
the gravitational anyon of Cho et al. [40] as an energy-spin composite. 

In the following we discuss only the most interesting case of
$\lambda \overline{l} < 0$.

Now the appropriate boundary conditions correspond to the
requirement that $f_{E,m}(r) $ is nonzero in either the interior
region ($r < r_0$) or in the exterior region ($r > r_0$). Thus
our boundary condition is
\begin{equation}
f_{E,m}(r_0) = 0
\end{equation}
The general solution of (7.6) is given by

\begin{equation}
f_{E,m}(r) = 
 Z_{ \overline{m}} \left( {\sqrt{E} \over \hbar} \, r \right)
\, ,
\end{equation}
where $Z_{\overline{m}}$ is an appropriate Bessel function of
order $\overline{m}$ (or a superposition of such functions).

\subsection{Interior Solutions ($r < r_0 $)}

The only Bessel functions not blowing up as $r\rightarrow 0$, 
{\it i.e.} giving a finite
probability $f\sp2\,r\,dr$ as $r\rightarrow 0$, and possessing positive zeros $r_0$ are those of the first kind of order
$\vert \bar{m}\vert $ with $E\ge 0$.

Then the possible eigenvalues $E_n(m)$ are determined by
\begin{equation}
J_{\bar{m}}\left[{\sqrt{E}\over \hbar}\left({\hbar\vert \lambda\bar{m}\vert\over 2\pi}\right)\sp{1\over 2}\right]\,=\,0
\label{result}
\end{equation}
with $\bar{m}$ given by (\ref{bar}).

Let us look in more detail at the case $\bar{m}> 0$, $\lambda<0$.
To simplify (\ref{result}) we define
\begin{equation}
\epsilon\,=\,{\vert \lambda\vert E\over 2\pi \hbar}.
\end{equation}
Thus (\ref{result}) takes the form
\begin{equation}
J_{\bar{m}}(\bar{m}\sp{1\over 2}\epsilon\sp{1\over2})\,=\,0.
\end{equation}

As $J_{\bar{m}}$, for fixed $\bar{m}>0$, has an infinite number of positive
zeros, which we denote by $y_n(\bar{m})$, $n=1,2..$ we see that due to (\ref{bar}), the eigenvalues $\epsilon_n(m)$ we are looking for are the positive fixed points of the equation
\begin{equation}
\epsilon\,=\,f_n(m+{1\over 2}\epsilon),
\label{neweq}
\end{equation}
where we have defined 
\begin{equation}
f_n(\bar{m})\,=\,{1\over \bar{m}}y_n\sp2(\bar{m}).
\end{equation}

The existence of positive fixed points $\epsilon$ of (\ref{neweq}) may be shown 
by using appropriate bounds for $f_n$ or $y_n$. For example for $n=1$ and $\bar{m}>0$ we have [41]

\begin{equation}
\bar{m}(\bar{m}+2)\,<\,\bar{m}f_1(\bar{m})\,<\,{4\over 3}(\bar{m}+1)(\bar{m}+5)
\end{equation}
and hence we have, with $\bar{m}=m+{1\over 2}\epsilon$ and for $m\in N$,
\begin{equation}
2(m+2)\,<\,\epsilon_1(m)\,<\,m+12\,+\,\left[(m+12)\sp2+8(m+1)
(m+5)\right]\sp{1\over2},
\end{equation}
{\it i.e.} we have proved the existence of $\epsilon_1(m)>0$ for $m=1,2..$
and give crude bounds for it. A better estimate for large $m$ can be obtained
by using the asymptotic formula [41]
$$y_1(\bar m)\,=\,\bar {m}\,+\,1.855757 \bar{m}\sp{1\over 3}\,+\,O(\bar{m}\sp{-{1\over 3}})$$
valid for large $\bar{m}$ leading to
\begin{equation}
\epsilon_1(m)\,=\,2m\,+\,9.35243m\sp{1\over 3}\,+\,O(m\sp{-{1\over 3}})
\end{equation}
valid for large positive $m$.

To get more insight the equation (\ref{neweq}) has to be solved numerically.
To do this we have to determine the behaviour of zeros of 
the Bessel function $J_k$ as a function of $k$.
Then  having determined this dependence we can find the values
of energy by the secant method.

 Luckily there are many
computer programs to determine zeros of Bessel functions
and in our work we have used the Maple program to 
perform this task. As the dependence 
of each zero is almost linear the numerical procedure of solving
(\ref{neweq}) converges rapidly.

In fig.1 we present the values of energy for $\epsilon\le 300$ as a function of
$m$. The plot looks like several curves; the lowest values correspond
to first zeros ({\it ie} $n=1$), the next ones to second zeros {\it ie} $n=2$ {\it etc}. The points lie so close that the figure may appear as a set of lines while,
in reality, we have here sets of points. The points appear to be 
(almost) equally spaced on each ``curve" - this is due to the 
approximate linearity of the positions of zeros of Bessel functions
as a function of $\bar{m}$. To check our values of energies we have also
solved (\ref{neweq}) differently; we approximated the positions
of zeros of the Bessel functions by a linear function and solved
the resultant equations for $\epsilon$. The obtained results were very similar
to what is shown in fig. 1 thus giving us confidence in our results.

Our results show that, for each value of $m$, there is a whole tower of
values of $\epsilon$ corresponding to different zeros of the Bessel functions.
The values of $\epsilon$ increase, approximately linearly, as we take
higher zeros ({\it ie} $y_n$ for larger $n$). The dependence on $m$ is only slightly more complicated;
for each order of the zero there is a value of $m$ for which the energy
is minimal and as we move away from this value the energy grows, approximately,
linearly. As $n$ increases the minimal values of $m$ increase, again, approximately linearly.

\begin{figure}[h]
\unitlength1cm
\hfil\begin{picture}(12,12)
\epsfxsize=11cm
\epsffile{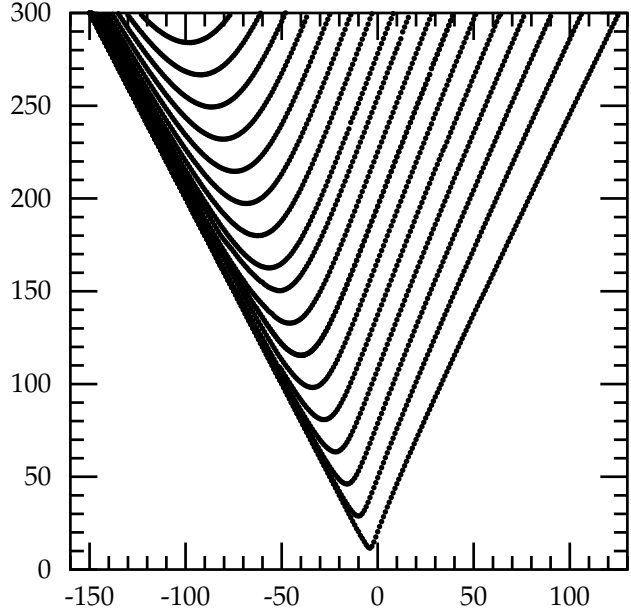}
\end{picture}
\caption{Energy as a function of $m$}
\end{figure}

Note that our numerical values of $\epsilon$ are consistent with the
asymptotic results mentioned before. Note also that for $\bar{m}<0$
and $\lambda>0$ the corresponding energy levels are obtained by changing the sign of $m$.

Summarising, we see that in the interior region $r<r_0$, where classical solutions are only possible for a finite time interval, we find quantum solutions which correspond to discrete bound states 
determined by the boundary condition at $r=r_0$.  Thus we see that this boundary
condition defines a geometric ``bag" for the quantum state.

\subsection{Exterior Solutions $(r>r_0)$}

First of all, let us note that there are no bound states solutions 
of (7.6) for $r>r_0$ as the only square integrable 
Bessel functions in $[r_0,\infty)$ are the
 modified
Bessel functions of the third kind, 
which have neither positive nor pure imaginary zeros (see {e.g.} [41]).

Scattering solutions are given by 
a superposition of Bessel functions of the first and second kind

\begin{equation}
f_{E,m}(r )\,=\,A_m(E)\,J_{\bar{m}}\left({\sqrt{E}\over \hbar}r\right)\,+\,
B_m(E)\,Y_{\bar{m}}\left({\sqrt{E}\over \hbar}r\right)
\end{equation}
with the ratio ${A_m\over B_m}$ determined by the boundary condition 
(7.8).
These solutions describe scattering
  on an obstruction of radius $r_0$, which is
dynamically determined.



\section{Two Charged Gravitationally Interacting 
 Particles in a Magnetic Field}

In this section we consider the motion considered in Sect.~6 and
7  of two particles, of equal
electric charge $e$, under the influence of an additional
static uniform magnetic field $B$
perpendicular to the plane of motion. We assume, 
for simplicity, that $eB>0$. 

\subsection{Classical Dynamics}

To have the Lagrangian $L$, describing the relative
  motion of two charged particles 
 in an additional constant magnetic field $B$ we add 
  to the Lagrangian considered in Section 6
 the term
\begin{equation}
L_{magn}\,=\,\frac{1}{4}\,e\,B\,\epsilon_{ij}\,x_i\,\dot x_j.
\ee
We obtain
\begin{equation}
L\,=\,(\xi_i\,+\,\frac{\lambda}{4\pi}\,\xi\sp2\,\partial_i\,\phi)\dot x_i\,-\,
\xi\sp2\,+\,\frac{1}{4}\,e\,B\,\epsilon_{ij}\,x_i\,\dot x_j
\end{equation}
and therefore, in accordance with (6.2) we have
\begin{equation}
H\,=\,\vec{\xi}\, \sp2\, .\end{equation}

The equations of motion for $x_i$, derived from (8.2), are the same as before, {\it ie}
\begin{equation}
\dot x_i\,=\,{2\xi_i\over 1+\frac{\lambda}{2\pi}\,\xi_l\,\partial_l\,\phi},
\end{equation}
where the covariant velocities
  $\xi_i$   for $B\ne 0$
      are not  conserved.
However, as $\xi_{i}$ satisfy
\be
\dot \xi_{i}\,=\,{eB\epsilon_{ij}\xi_j\over 1+\frac{\lambda}{2\pi}\xi_l\partial_l\phi}
\ee
we see (from (8.4-5)) that the conserved quantities are now
\be
\tilde \xi_i:\,=\,\xi_{i}\,-\,\frac{eB}{2}\,\epsilon_{ij}\,x_j.
\ee

From the equations of motion (8.4-5) and the Hamiltonian (8.3) we derive
the following, nonstandard, symplectic structure for the noncanonical set of variables $(\vec x, 
\vec\xi)$ or $(\vec x,\vec{\tilde{\xi}})$, respectively:
\setcounter{equation}{0}
\renewcommand{\theequation}{\thesection.7\alph{equation}}

\be
\{x_i,\,x_j\}\,=\,0
\ee

\be
\{x_i,\,\xi_j\}\,=\,\{x_i,\,\tilde\xi_j\}\,=\,\delta_{ij}\,-\,{\frac{\lambda}{2\pi} \,\xi_i\partial
_j\phi\over 1+\frac{\lambda}{2\pi}\xi_l\partial_l\phi}
\ee
\be
\{\xi_i,\,\xi_j\}\,=\,{\frac{1}{2}eB\epsilon_{ij}\over 1+\frac{\lambda}{2\pi}\xi_l\partial_l\phi}
\ee
but
\be
\{\tilde\xi_i,\,\tilde\xi_j\}\,=  - \,\frac{1}{2}eB\epsilon_{ij}.
\ee

\setcounter{equation}{7}
\renewcommand{\theequation}{\thesection.\arabic{equation}}
It is easy to check that the Poisson brackets (8.7a-d) satisfy the Jacobi identities.

With the canonical momentum $p_i$ derived from (8.2)
\be
p_i\,=\,\xi_i\,+\,\frac{\lambda}{4\pi}\,H\,\partial_i\phi\,-\,\frac{eB}{4}\,\epsilon_{ij}\,x_j
\ee
we find that the relation between  the two conserved angular
  momenta $l, \overline{l}$ is not modified
\be 
l:\,=\vec x\wedge\vec p\,=\,\bar{l}\,+\,\frac{\lambda}{4\pi}H;
\ee
however,  $\bar{l}$ is given by
\be
\bar{l}:\,=\,\vec x\wedge \vec \xi\,+\,\frac{1}{4}\,e\,B\,r\sp2
\, .
\ee
We note that due to (8.9-10) the form of the denominator in the 
equation of motion (8.4-5) is different when compared with the 
$B=0$ case:
\be
1\,+\,\frac{\lambda}{2\pi}\,\xi_l\partial_l\phi\,=\,\left(1-\frac{\lambda}{8\pi}
eB\right)\,+\,{\lambda\bar{l}\over 2\pi r\sp2}.
\ee

Let us consider just the nonconfined motion, {\it ie} (8.11) 
should be nonvanishing for all $0<r<\infty$.  When $B=0$ this would 
correspond 
to the case of $\lambda \bar{l}>0$ - {\it ie} ``the less 
interesting case" of the previous section.

Now, however,  we have to
distinguish two subcases:

Either
\setcounter{equation}{0}
\renewcommand{\theequation}{\thesection.12\alph{equation}}
\be
\mbox{A}:\qquad 1-\frac{\lambda}{8\pi}\,eB\,>\,0,
\ee
which allows both signs of $\lambda$,
\be
\mbox{and}\quad \lambda\bar{l}\,>\,0
\ee

\setcounter{equation}{12}
\renewcommand{\theequation}{\thesection.\arabic{equation}}
or
\setcounter{equation}{0}
\renewcommand{\theequation}{\thesection.13\alph{equation}}
\be
\mbox{B}:\qquad 1-\frac{\lambda}{8\pi}\,eB\,<\,0,
\ee
which allows only $\lambda>0$
\be
\mbox{and}\quad \bar{l}\,<\,0.
\ee

\setcounter{equation}{13}
\renewcommand{\theequation}{\thesection.\arabic{equation}}

Note that when $B\ne0$ we can have a nonconfined motion also
for $\lambda\bar{l}<0$.

In analogy to the free field case we introduce for
\be
{\lambda \bar{l}\over 1-{\lambda\over 8\pi}eB}\,<\,0
\ee
the boundary $r_0$ between the interior ($r<r_0$) and the exterior
 ($r>r_0$) region defined by
\be
r_0\sp2\,=\,{1\over 2\pi}\left\vert {\lambda \bar{l}\over 1-{\lambda\over 8\pi}eB}\right\vert.
\ee
When we consider the confined motion we consider the motion in the
interior region ($r<r_0$). In this case we have to consider two 
further subclasses:

Either
\setcounter{equation}{0}
\renewcommand{\theequation}{\thesection.16\alph{equation}}
\be
\mbox{C}:\qquad 1-\frac{\lambda}{8\pi}\,eB\,>\,0,
\ee
which allows both signs of $\lambda$,
\be
\mbox{and}\quad \lambda\bar{l}\,<\,0
\ee

\setcounter{equation}{16}
\renewcommand{\theequation}{\thesection.\arabic{equation}}
or
\setcounter{equation}{0}
\renewcommand{\theequation}{\thesection.17\alph{equation}}
\be
\mbox{D}:\qquad 1-\frac{\lambda}{8\pi}\,eB\,<\,0,
\ee
which allows only $\lambda>0$
\be
\mbox{and}\quad \bar{l}\,>\,0.
\ee

\setcounter{equation}{17}
\renewcommand{\theequation}{\thesection.\arabic{equation}}

Note that when $B\ne 0$ we can have a confined motion also for $\lambda\bar{l}>0$.

Comparing all four subcases we conclude that we obtain 
confinement $\leftrightarrow$ nonconfinement transitions by tuning the
strength of the magnetic field $B$. To be more specific let us 
consider {\it eg} $\lambda>0$ and $ \bar{l}>0$. Then we obtain, for $eB$
increasing from low values towards ${8\pi\over \lambda}$, a transition from case $A$ to case $D$ {\it ie} a nonconfinement $\rightarrow$ confinement
transition.  This transition is continuous because we have $r_0=\infty$ at
the transition point.

In order to quantize our system we have to 
generalize for  $B\ne 0$ the relation (7.2a).
We start with (8.8) rewritten as
\be
\xi_i\,=\,p_i\,-\,\frac{\lambda}{4\pi}\,H\,\partial_i\,\phi\,+\,\frac{eB}{4}\,\epsilon_{ij}\,x_j,
\ee
square it and then using (8.3) and (8.9-10) we find for the Hamiltonian $H$
\be
H\,=\,p\sp2\,-\,\frac{l\sp2}{r\sp2}\,+\,\frac{\bar{l}\sp2}{r\sp2}\,+\,\frac{1}{16}(eB)\sp2 r\sp2\,-
\,\frac{1}{2}\,eB\bar{l}.
\ee

\subsection{Quantization}

Following the 
 method presented
 in Section 7.1 we obtain 
from (8.19) the following nonstandard Schr\"odinger equation for the radial 
wave function $f_{E,m}$
\be
\left[-\hbar\sp2\left(\partial_r\sp2\,+\,\frac{1}{r}\partial_r\,-\,{\bar{m}\sp2
\over r\sp2}\right)\,+\,\left(\frac{eB}{4}\right)\sp2r\sp2\,-\,E\,-\frac{\hbar}{2}eB\bar{m}\right]\,f_{
E,m}(r)\,=\,0,
\ee
which generalizes (7.6). Like in the previous case, the noninteger eigenvalue
$\bar{m}$ is related to the integer $m$ and energy $E$ by the eq. (7.7).

\subsubsection{Nonconfined motion}

The eigenvalue problem (8.20) is identical to the two anyon problem
in a constant $B$ field  with the statistics
parameter being proportional to the energy value $E$. So, for the energy levels
$E_{n,m}$ we obtain ([42-44]) 
\be
E_{n,m}\,=\,\hbar\,e\,B\,(n\,+\,\frac{1}{2}\vert \bar m\vert\,+\,\frac{1}{2})
\,-\,\frac{\hbar}{2}eB\bar{m},
\ee
with $n=0,1,2,..$ 
Thus, according to the two cases A and B defined
 above and for different signs of $\lambda$ 
we have to distinguish between three cases:

{\bf Case A with $\lambda>0$}

Due to (8.12b) and (7.7) we have
\be
\bar m\,=\,m\,-\,\frac{\lambda}{4\pi\hbar}\,E_{n,m}\,>\,0.
\ee
Therefore we get from (8.21)
\be 
E_{n,m}\,=\,\hbar e B(n+\frac{1}{2})
\ee
corresponding to the case II in \cite{41}.

Combining (8.22) and (8.23) gives us
\be
m\,>\,{\lambda eB\over 4\pi}\left(n\,+\,\frac{1}{2}\right),
\ee
where, due to (8.12a), the product of the coupling strength $\lambda$ and of the 
field strength $B$ is bounded
\be 
\lambda e B\,<\,8\pi.
\ee

{\bf Case B with $\lambda>0$}

Due to (8.13b) and (7.7) we have
\be
\bar m\,=\,m\,-\,\frac{\lambda}{4\pi\hbar}\,E_{n,m}\,<\,0.
\ee
Thus, from (8.21) we obtain
\be
E_{n,m}\,=\,\hbar eB\left(n\,+
 \,\vert \bar m\vert     \, + \,\frac{1}{2}\right),
\ee
which corresponds to the case I in \cite{41}.

Inserting (8.26) into (8.27) we obtain
\be
E_{n,m}\,=\,{\hbar eB\over 1-\frac{\lambda eB}{4\pi}}\left(n-m+\frac{1}{2}\right).
\ee

Note that in order to satisfy (8.13a) the product of the coupling
strength $\lambda$ and of the field strength $B$ must be above the mininum value
\be
\lambda e B\,>\,8\pi.
\ee

Therefore we obtain from (8.26) and (8.28)
\be 
m\,>\,\frac{\lambda eB}{4\pi}\left(n\,+\,\frac{1}{2}\right).
\ee

Thus we see that $E_{n,m}$ is bounded from below by
\be
E_{n,m}\,>\,\hbar e B(n+\frac{1}{2}).
\ee

{\bf Case A with $\lambda<0$}

The condition (8.12a) is now fulfilled automatically. From (8.12b) we obtain 
(8.26) and therefore we have the result identical to (8.28).

Now we obtain
\be
m\,<\,-{\vert \lambda \vert eB\over 4\pi}(n+\frac{1}{2})
\ee
and $E_{n,m}$ is bounded from below again by
\be
E_{n,m}\,>\,\hbar eB(n+\frac{1}{2}).
\ee
\subsubsection{Confined motion}

Now we have to consider a nonstandard Schr\"odinger equation (8.20) in the
interior region ($r<r_0$) with the boundary condition as in the previous 
case
\be
f_{E,m}(r_0)\,=\,0.
\ee

The regular solution of (8.20) is given, up to a normalisation factor, by
[45]
\be
f_{E,m}(r)\,=\,r\sp{\vert \bar{m}\vert}\,e\sp{-{\beta r\sp2\over 2}}
\phi\left({\vert \bar{m}\vert -\bar{m}+1\over 2}-\gamma,1+\vert \bar{m}\vert;\beta r\sp2\right)
\ee
with

\setcounter{equation}{0}
\renewcommand{\theequation}{\thesection.36\alph{equation}}
\be
\beta\,=\,{eB\over 4\hbar}
\ee
and
\be
\gamma\,=\,{E\over \hbar eB}
\ee
and where $\phi(a,b;z)$ denotes the confluent hypergeometric function.
\setcounter{equation}{36}
\renewcommand{\theequation}{\thesection.\arabic{equation}}

From (8.15), (8.34) and (8.35) we conclude that our energy levels
$E_{n,m}$ are determined by the roots of the equation, which only
be solved numerically,
\be
\phi\left({\vert \bar{m}\vert -\bar{m}+1\over 2}-{E\over \hbar eB},1+\vert \bar{m}\vert;{eB\over 8\pi}\left\vert{\lambda \bar{m}\over 1-{\lambda\over 8\pi}eB}\right\vert \right)\,=\,0,
\ee
where $\bar{m}$ is a function of $E$ as given by (7.7).

{\bf The zero $B$ field limit}

From the well known relation [45]
\be
\lim_{a\rightarrow \infty}\,\phi(a,b;-{x\over a})\,=\,\Gamma(b)x\sp{{1\over 2}
(1-b)} \,J_{b-1}(2\sqrt{x})
\ee
we obtain, from (8.35) and (8.37),  in the vanishing $B$ limit,
the considered in section 7, respectively, the wave function and the eigenvalue condition.

{\bf The high $B$ field limit}

Without going into the numerics we can conclude from (8.37) that for $eB\rightarrow\infty$ the energy levels
increase with the increase of $eB$, at least, linearly. We prove this statement
by assuming the contrary. Then one has at the l.h.s. of (8.37), with $\bar{m}>0$,
due to (8.16-17), $\phi(\frac{1}{2}, 1+\bar{m}; \bar{m})$ for which we have the inequality
\be 
\phi(\frac{1}{2}, 1+\bar{m}; \bar{m})\,\ge\,1\quad \mbox{for}\quad \bar{m}>0
\ee
in contradistinction to (8.37).
\vfil
\eject
{\bf Confinement $\rightarrow$ Nonconfinement transitions}

According to (8.15) confinement $\rightarrow$ nonconfinement transitions
occur if
$r_0\rightarrow \infty$, {\it ie} if 
\be
1\,-\,{\lambda\over 8\pi}eB\,\rightarrow\,0.
\ee
But from the asymptotic behaviour of the confluent 
hypergeometric function [45]
\be
\phi(a,b;x)\,\rightarrow_{x\rightarrow \infty}\,{\Gamma(b)\over \Gamma(a)}
\,e\sp{x}\,x\sp{a-b}
\ee
we conclude that the r.h.s. vanishes only if $a=0,-1,-2,...$, {\it ie}
for (8.37), if the energy levels are given by (8.21) as required.

\section{Conclusions and Outlook}
The aim of this paper has been to consider, for two--dimensional
nonrelativistic particles, a  new interaction scheme, generated by
the coupling to the topological torsion Lagrangian. We have shown that
only in $D=2+1$ one can write  the torsion Lagrangian in the
form of a bilinear Chern--Simons term, described by the
translational gauge fields  (threebeins) multiplied by their
Abelian field strength (torsion fields). By a suitable choice of
the reparametrization gauge, we have then determined the solutions 
of  the two--body dynamics
with fractional  angular momentum eigenvalues corresponding to trajectories confined   to finite regions of $d=2$ space. The
quantization problem was described by a new type of Schr\"{o}dinger
equation, with a second order time derivative, which in the stationary
case gave us a nonlinear energy eigenvalue problem, describing
infinite sequence of bound states. 

The eigenvalues of energy were determined numerically and we showed
that the spectrum was descrete and characterised by two
integers; one of them corresponding to the rotational integer quantum 
number $m$ and the other, another integer - $n$, described
which zero of the Bessel function the state corresponded to.
The dependence on both of these integer-value parameters
was approximately linear - as we discussed this in section 7.

We have also noted that due  to the  topological  form of the
free gravitational field action the  case of two dimensions is
exceptional. In $d=3$ one can also look      for a
modification of the standard gravitational interactions ({e.g.}
 of the Newton potential in the lowest order static approximation) by 
 adding to the Hilbert--Einstein 
  action appropriate bilinear torsion terms (see e.g. [7]).
However,  the confinement due to the presence of
torsion in $D=3+1$ gravity leads in the three--dimensional space to
an additional potential term of a harmonic form
  (see e.g. [8]).
Such a mechanism of confinement is different from our proposal, where
the two-particle Lagrangian becomes singular at the boundary
$r_0$ of a dynamically determined compact region of space.
Equivalently, confinement is obtained from the singularity of the nonstandard 
symplectic structure within the Hamiltonian formalism.

Our considerations have thus shown us [2,3,13] that the geometric bag
formation\footnote{In the recent literature
one finds geometric bag models determined by the
singularity of a given metric (cp. [46] and the literature quoted therein)} based on our dynamical assumptions are possible in $d=1$
and $d=2$ spaces, but it is not clear how to  obtain analogous
solutions, to be generated by nonstandard  gravitational interactions,
in the $d=3$ case, {\it i.e.} in the standard physical space--time. 
It is worth pointing out, however, that our geometric bag solutions carry some
resemblance to the fields describing black holes (see {e.g.}
the dynamical division of space into two disconnected
 domains -- interior and exterior).
At present we can only hope that all such consequences for $d > 2$ will be
 further clarified in the future development of our approach.

\subsection*{Acknowledgment}
The  authors  would like to thank F.W. Hehl for his interest in 
our work and correspondence.
One of the authors (WJZ) would like to thank J.F. Blowey and
J.P. Coleman for their help with the numerics described in Section 7.

\end{document}